\documentclass[10pt, aps, prd, superscriptaddress, nofootinbib, showpacs, twocolumn]{revtex4-2}
\usepackage[english]{babel}
\usepackage{amsmath, amsfonts, amsthm, amssymb, mathrsfs}
\usepackage[all]{xy}
\usepackage{bm}
\usepackage{stmaryrd}
\usepackage{graphicx}
\usepackage[colorlinks]{hyperref}
\usepackage{color}
\usepackage{subcaption}
\usepackage{amscd}

\usepackage{pgfplots, grffile, tikz}
\pgfplotsset{compat=newest}
\usetikzlibrary{plotmarks, arrows.meta}
\usepgfplotslibrary{patchplots}

\usepackage{dsfont}

\makeatletter

\renewcommand*{\p@subsection}{}

\renewcommand*{\p@subsubsection}{}
\makeatother
\numberwithin{equation}{section}

\DeclareMathOperator*{\Res}{Res}
\renewcommand{\Re}{\mathop{\mathrm{Re}}\nolimits}
\renewcommand{\Im}{\mathop{\mathrm{Im}}\nolimits}

\newcommand{\calA}{\mathcal A}
\newcommand{\calE}{\mathcal E}
\newcommand{\calG}{\mathcal G}
\newcommand{\calH}{\mathcal H}
\newcommand{\calK}{\mathcal K}
\newcommand{\calM}{\mathcal M}

\newcommand{\bbB}{\mathbb B}
\newcommand{\bbC}{\mathbb C}
\newcommand{\bbW}{\mathbb W}
\newcommand{\bbL}{\mathbb L}
\newcommand{\bbM}{\mathbb M}
\newcommand{\bbR}{\mathbb R}
\newcommand{\bbZ}{\mathbb Z}

\newcommand{\frakL}{\mathfrak{L}}

\newcommand{\lp}{\left(}
\newcommand{\rp}{\right)}
\newcommand{\lb}{\left[}
\newcommand{\rb}{\right]}
\newcommand{\lc}{\left\{}
\newcommand{\rc}{\right\}}
\newcommand\tln[2]{\substack{#1\\#2}}

\renewcommand{\a}{\alpha}
\newcommand{\g}{\gamma}
\newcommand{\m}{\mu}
\newcommand{\n}{\nu}
\newcommand{\s}{\sigma}

\newcommand{\pp}{\partial}
\newcommand{\nb}{\nabla}

\newcommand{\h}{\hat}

\begin{document}

\title{Multiple Mellin-Barnes integrals in Schwinger-DeWitt technique}

\author{A. O. Barvinsky}
\email{barvin@td.lpi.ru}
\affiliation{Theory Department, Lebedev Physics Institute, Leninsky Prospect 53, Moscow 119991, Russia}
\affiliation{Institute for Theoretical and Mathematical Physics, Moscow State University, Leninskie Gory, GSP-1, Moscow, 119991, Russia}

\author{A. E. Kalugin}
\email{kalugin.ae@phystech.edu}
\affiliation{Theory Department, Lebedev Physics Institute, Leninsky Prospect 53, Moscow 119991, Russia}

\author{W. Wachowski}
\email{vladvakh@gmail.com}
\affiliation{Theory Department, Lebedev Physics Institute, Leninsky Prospect 53, Moscow 119991, Russia}

\begin{abstract}
We consider off-diagonal asymptotic series for integral kernels of functions of Laplace-type operators on curved backgrounds. These expansions are obtained by applying integral transforms to the DeWitt series for the heat kernel of the corresponding operator and thus represent a DeWitt-type series in the heat kernel coefficients and some hypergeometric-type functions of the Synge world function (which we call basis kernels). Basis kernels of certain class of operator functions were found previously in terms of $N$-fold Mellin-Barnes integrals. In this paper we study series representations of the corresponding Mellin-Barnes integrals in both non-resonant and resonant cases and suggest a physical interpretation for the emerging series, which is related to the UV and IR properties of operator functions.
\end{abstract}

\maketitle
\begingroup
  \hypersetup{linkcolor=blue}
  \tableofcontents
\endgroup

\section{Introduction} \label{sec:introduction}

This paper is a sequel to \cite{BKWletter, BKW25a} and is devoted to the study of several multiple Mellin-Barnes (MB) integrals arising in integral kernels expansions of functions of differential operators within the Schwinger-DeWitt technique. In this sense, one could say that it is of a highly technical nature. However, since it presents several simple yet important examples of multiple MB integrals, we believe it may also be useful to anyone encountering these special functions. Therefore, before proceeding directly to particular task at hand, we provide a comment on its importance.

$N$-fold MB integrals $H(\bm{\alpha}, A, \bm{a} | \bm{z})$ are functions of $N$ complex arguments $\bm{z}$, parametrically determined by two vectors $\bm{\alpha}$ and $\bm{a}$, and some matrix $A$. We will give their precise definition a little later, in Sec.~\ref{DefSubSec}. For now we note that they are generalizations of both the Fox $H$-functions \cite{Fox1961, Braaksma, Slater1966, Marichev1983, Paris2001} and the solutions of Gelfand-Kapranov-Zelevinsky (GKZ) systems \cite{GelfandPapers, Gelfand1990, Gelfand1994}.\footnote{$N$-fold MB integrals are ``fractional'' generalizations of $A$-hypergeometric functions of GKZ in the same sense in which the Fox $H$-functions generalize the usual hypergeometric functions ${}_pF_q$.} They arise in a wide variety of applications. To fully appreciate their enormous significance for physics, it is enough to note that all scalar Feynman integrals can be expressed in terms of multiple MB integrals, where the parameters $\bm{\alpha}$ and $A$ are determined by the topology of the Feynman diagram, the parameters $\bm{a}$ are determined by the spacetime dimension and the parameters of analytic regularization, and the arguments $\bm{z}$ are dimensionless combinations of masses and external momenta \cite{Weinzierl2022, Dubovyk2022, Klausen2023}.

However, despite their fundamental role, multiple MB integrals remain insufficiently studied. In the literature, one can find only two series of works devoted to the systematic study of their properties: these are the purely mathematical papers of A. K. Tsikh \emph{et al.} in the 1990s \cite{Tsikh1992, Passare1994, Passare1996, Zhdanov1998} and, in recent years, the papers of S. Friot \emph{et al.} \cite{Friot2012, Ananthanarayan2021, Banik2023}, who dealt with them from the point of view of symbolic computations in the context of studying Feynman integrals.

The key task in studying multiple MB integrals is to obtain their representations in the form of Horn series (power series in $\bm{z}$) or its generalization (containing factors logarithmic in $\bm{z}$). Generally speaking, there may be a lot of such representations, and the corresponding series may be either convergent or merely asymptotic. Accordingly, in addition to obtaining the series representations themselves, it is also necessary to be able to find their regions of convergence (specific to each series), or in what region of variables $\bm{z}$ the series is asymptotic. This problem will be considered in detail below in Subsec.~\ref{SeriesSubSec}. After that the general statements obtained will be applied to a number of specific examples that arose in \cite{BKWletter, BKW25a}. However, it should be emphasized that the same methods can be applied with equal success to other multiple MB integrals arising in problems that may differ greatly from ours.

The examples of multiple MB integrals that we will consider further in Secs.~\ref{NonresSec} and \ref{ResonantSec} will be taken without derivation from our paper \cite{BKW25a}. However, it is appropriate to at least briefly recall the problem in which they arise. Therefore, in the remainder of this Introduction, we will repeat the main ideas of our method, a much more detailed exposition of which can be found in papers~\cite{BKWletter, BKW25a}.

Consider a spacetime $\calM$---$d$-dimensional Riemannian manifold with metric $g_{ab}$ and Levi-Civita connection $\nabla_a$. Let $\h F(\nb)$ be a Laplace-type operator (i.e., minimal of 2nd order) acting on sections of a vector bundle over $\calM$:
\begin{equation}
\hat F(\nabla) = \hat 1\Box + \hat P,
\end{equation}
where $\Box = -g^{ab}\nabla_a\nabla_b$ is the covariant Laplacian, hats denote the matrix structure in the fibers, and $\hat P=\h P(x)$ is an arbitrary matrix (a ``potential term''). The heat kernel for $\hat F(\nabla)$ is defined as the integral kernel of the operator exponential:
\begin{equation}
\hat K_F(\tau | x, x') = \exp(-\tau\hat F)\, \delta(x, x'),
\end{equation}
where $\delta(x, x')$ is the invariant delta function. The heat kernel is a two-point (i.e., depending on two points $x,x'\in\calM$, as well as on the additional proper time parameter $\tau$) matrix-valued function. A remarkable fact \cite{DeWitt1965, Barvinsky1985} is that there is an asymptotic DeWitt's expansion of the heat kernel in the ultraviolet (UV) limit $\tau\to0$
\begin{align}
&\hat K_F(\tau |\, x,x') = \sum\limits_{n=0}^\infty\!B_{\frac{d}2 - n}(\tau,\sigma) \cdot \hat a_n[F | x,x'], \label{HeatKernelExpansion} \\
&B_\alpha(\tau,\sigma)= \frac{\tau^{-\alpha}}{(4\pi)^{d/2}}\! \exp\left(-\frac{\sigma}{2\tau}\right), \label{InitialKernel}
\end{align}
where $\sigma = \sigma(x, x')$ is the Synge world function, and $\hat a_m[F | x,x']$ are off-diagonal heat kernel (or HaMiDeW) coefficients.\footnote{We include the quasiclassical Pauli-Van Vleck-Morette determinant $\Delta^{1/2}(x, x')$, which is usually written explicitly, in coefficients $\hat a_m[F | x,x']$.} Coincidence limits $\nabla_{b_1}\cdots\nabla_{b_k} \hat a_n \big|_{x=x'}$ are local invariants of mass dimension $2n+k$ that can be recursively computed up to the required order as combinations of $\hat P$, the Riemann tensor, the curvature in the bundle, and covariant derivatives thereof \cite{DeWitt1965, Barvinsky1985, Gilkey1975, Gilkey1995}. They play a fundamental role in quantum field theory (QFT) in curved spacetime, since they determine the one-loop effective action of the theory with the wave operator $\hat F(\nabla)$.

However, in applications (in particular, for studying theories with non-minimal wave operators \cite{BarvinskyKalugin2024, Barvinsky25}), it is often necessary to work not only with the operator exponential $\exp(-\tau\hat F)$, but also with operator functions $f(\hat F)$ of a more complex form, for example, $\exp(-\tau\hat F)/\hat F^\mu$. Therefore, we would like to obtain a representation for their integral kernels,
\begin{equation}
\hat K\big[f(F) \big| x,x' \big] = f(\hat F)\, \delta(x,x'),
\end{equation}
as a functional series, similar to \eqref{HeatKernelExpansion}-\eqref{InitialKernel}. To do this, it suffices to note that any operator function $f(\hat F)$ can be written as an integral transform of the operator exponential $\exp(-\tau\hat F)$
\begin{align}
&f(\hat F) = \frakL_f\, e^{-\tau\hat F} = \int\limits_0^\infty d\tau\, f^*(\tau)\,e^{-\tau\hat F}, \label{frakLtransformEq} \\
&f^*(\tau) = \int\limits_C \frac{d\lambda}{2\pi i}\; f(\lambda)\, e^{\tau\lambda}, \label{InvLaplaceTransform1}
\end{align}
which should be understood as the Laplace transform from the number variable $\tau$ to the operator variable $\hat F$. Now we apply the integral transform $\frakL_f$ term-by-term to the DeWitt expansion \eqref{HeatKernelExpansion}, and obtain the desired expansion
\begin{align} 
&\hat K\big[f(F) \big| x,x' \big] \simeq \sum\limits_{n=0}^\infty \bbB_{\frac{d}{2}-n}[f | \sigma] \cdot \hat a_n[F | x,x'], \label{Bexpansion} \\
&\bbB_\alpha[f | \sigma] = \frakL_f B_\alpha(\tau,\sigma) = \int\limits_0^\infty d\tau\, f^*(\tau)\,B_\alpha(\tau,\sigma). \label{defLf}
\end{align}

The transforms we performed are possible because we work outside the coincidence limit $x=x'$, thus avoiding the singularities that arise it it. \emph{Off-diagonality}, which is essentially a form of regularization by separating points, is a key ingredient of our approach, allowing us to flexibly use the powerful apparatus of integral transforms.

A remarkable property of the expansion \eqref{Bexpansion} is that it explicitly distinguishes the data of two types: whatever the function $f$, information about it is contained in the scalar functions $\bbB_\alpha[f | \sigma]$ \eqref{defLf}, which we call \emph{basis kernels}. And all the information about the geometry of the spacetime $\calM$ and the vector bundle, as well as the specific form of the operator $\hat F(\nabla)$, is still encoded in the familiar HaMiDeW coefficients $\hat a_n[F | x,x']$. This property is an off-diagonal generalization of the phenomenon called "functoriality" in the literature \cite{Gilkey1975, Gilkey1995}, so we call it \emph{off-diagonal functoriality}.

An important point, crucial for our strategy, is the following: as we explain in detail in \cite{BKWletter, BKW25a}, since the expansion \eqref{Bexpansion} is obtained by term-by-term integration of the DeWitt expansion \eqref{HeatKernelExpansion}, which is valid only in the UV limit $\tau\to0$, it is not in fact a total expansion of the integral kernel $\hat K\big[f(F) \big| x,x' \big]$, but only its UV half (which we express by using the $\simeq$ sign instead of equality in \eqref{Bexpansion}). To obtain the second, infrared (IR) half of the total expansion, we would have to similarly integrate term-by-term the asymptotic expansion of the heat kernel $\hat K_F(\tau | x, x')$ in the opposite IR limit $\tau\to\infty$ (finding which, however, is a much more difficult problem).

In other words, the same point can be expressed as follows: as is well known, according to the Fubini-Tonelli theorem, the order of summation and integration can be changed only if all intermediate sums and integrals converge absolutely. In our case, neither one nor the other holds: the DeWitt expansion \eqref{HeatKernelExpansion} is an asymptotic (divergent) series, but in addition, the integral in \eqref{defLf} for sufficiently small $\Re\alpha$ can diverge at the IR limit $\tau=\infty$, and therefore requires some additional regularization. The expressions \eqref{Bexpansion}-\eqref{defLf} imply regularization by means of analytic continuation: if the integral converges in a certain region of its parameters, then the analytic continuation of the resulting expression beyond this region yields the regularized value of the divergent integral.

But we can do in another way: add a massive term to the operator $\hat F \mapsto \hat F + m^2$, and take it into account ``nonperturbatively'' in the prefactor. This means that in the DeWitt expansion \eqref{HeatKernelExpansion}, the function $B_\alpha(\tau, \sigma)$ \eqref{InitialKernel} should be replaced by the expression
\begin{equation} \label{IRregKernel}
W_\alpha(\tau, \sigma, m^2) = \frac{\tau^{-\alpha}}{(4\pi)^{d/2}} \exp\left(-\frac{\sigma}{2\tau} - \tau m^2 \right).
\end{equation}
After this, the integral transform $\frakL_f$ \eqref{frakLtransformEq} acting term-by-term leads to the following modified expansion
\begin{equation}
\hat K\big[f(F+m^2) \big| x,x' \big] \simeq \sum\limits_{n=0}^\infty \bbW_{\frac{d}{2}-n}[f | \sigma, m^2] \cdot \hat a_n[F | x,x']. \label{Wexpansion}
\end{equation}
It is in a new family of functions
\begin{equation} \label{BHtransform}
\bbW_\alpha[f | \sigma, m^2] = \frakL_f W_\alpha(\tau, \sigma, m^2),
\end{equation}
which we call \emph{complete massive kernels}. Note that the presence of an exponentially decaying factor $\exp(-\tau m^2)$ in $W_\alpha(\tau, \sigma, m^2)$ \eqref{IRregKernel} ensures good convergence of the integral in \eqref{BHtransform} at the IR limit $\tau=\infty$. Accordingly, the functions $\bbW_\alpha[f | \sigma, m^2]$ are well defined for all values of the parameter $\alpha$.

In \cite{BKW25a} we used a simple but effective technique of integral transforms to compute the basis $\bbB_\alpha[f|\sigma]$ and complete massive $\bbW_\alpha[f | \sigma, m^2]$ kernels for operator functions $f(\hat F)$ of three types important for applications, namely for $\exp(-\tau\hat F^\nu)/\hat F^\mu$, $1/(\hat F^\mu + \lambda)$, and finally for $\exp(-\tau\hat F^\nu)/(\hat F^\mu + \lambda)$. It turned out that for these operator functions (as well as for many others of interest), the corresponding kernels can be universally represented as $N$-fold MB integrals, where $N$ is one less than the number of dimensional parameters in the problem under consideration.

In this paper, which is a continuation of \cite{BKW25a}, we take the MB integrals obtained there as a starting point and examine their properties in detail in both non-resonant and resonant cases in Secs.~\ref{NonresSec} and \ref{ResonantSec}, respectively. Before discussing these specific examples, however, we need to examine the general properties of multiple MB integrals in greater detail, which will be done in the preliminary Sec.~\ref{NfoldMBsec}.

\section{Elements of MB integral theory} \label{NfoldMBsec}

Multiple MB integrals will be defined as a multiple inverse Mellin transform. In this regard, it is useful to recall some properties of Mellin transforms, which will be done in Subsec.~\ref{MellinTransSec}. Next, in Subsec.~\ref{DefSubSec}, we give the precise definition and discuss some subtle questions about it. Finally, Subsec.~\ref{SeriesSubSec} will be devoted to series representations of MB integrals.

\subsection{Mellin transforms} \label{MellinTransSec}

Let some function $H(z)$ be absolutely integrable on any interval that is a proper subset of the positive semi-axis $(0, +\infty)$, and $H(z) \sim z^{-a}$ for $z\to0$, and $H(z) \sim z^{-b}$ for $z\to\infty$ (where $a, b$ can be equal to $\pm\infty$ which means that $H(z)$ grows or decreases faster than any power). Then, obviously, the integral
\begin{equation} \label{DirectMellinDef}
h(s) = \int\limits_0^\infty dz\,z^{s-1} H(z),
\end{equation}
converges at the lower limit $z=0$ for any complex parameter $\Re s>a$ and at the upper limit $z=\infty$ for any $\Re s<b$. Consequently, if $a<b$, the integral \eqref{DirectMellinDef} converges absolutely for any $s$ lying in a vertical strip $a < \Re s < b$ , which is called the \emph{fundamental strip} of the function $H(z)$. Then the integral \eqref{DirectMellinDef} defines a holomorphic function $h(s)$ on it, which can be analytically continued beyond the fundamental strip and is called the \emph{direct Mellin transform} of the function $H(z)$.

Then the Mellin inversion theorem states that the integral
\begin{equation} \label{InverseMellinDef}
H(z) = \int\limits_{w-i\infty}^{w+i\infty} \frac{ds}{2\pi i}\, z^{-s} h(s),
\end{equation}
where the integration is over a vertical straight contour lying in the fundamental strip $a<w<b$, recovers the original function $H(z)$.

Here it is necessary to note two very important circumstances. First, in contrast to the Fourier transforms more familiar to physicists, the direct and inverse Mellin transforms are fundamentally asymmetric. The direct Mellin transform \eqref{DirectMellinDef}, when it is defined at all, is always unique: for each function $H(z)$ its fundamental strip and (if it is not empty) the Mellin image $h(s)$ in it are uniquely defined. On the contrary, its inverse \eqref{InverseMellinDef} is ambiguous: if we are given some function $h(s)$ that has several vertical analyticity strips $a_k < \Re s < b_k$ (separated by some singularities), then the integrals \eqref{InverseMellinDef} in which the vertical contours lie in these strips will define different functions $H_k(z)$ (differing from each other by residues in the singularities of the function $h(s)$).

Let us illustrate this with a simple example: consider the usual exponential $H(z) = e^{-z}$. We can see that the integral \eqref{DirectMellinDef} for it exactly coincides with \eqref{GammaDef}, whence it follows that the fundamental strip coincides with the half-plane $\Re s>0$, and the Mellin image is the gamma function $h(s) = \Gamma(s)$. However, the gamma function $\Gamma(s)$, in addition to the half-plane $\Re s>0$, has an infinite number of other analyticity strips $-n < \Re s < -n+1$. And if we calculate the integral \eqref{InverseMellinDef} with a contour lying in such a strip, then obviously the result will differ by residues at the $n$ poles of the function $\Gamma(s)$ lying to the right of this contour. The result is the function obtained from the exponential by subtracting the first $n$ terms of its Taylor series
\begin{equation}
H_n(z) = e^{-z} - \sum\limits_{k=0}^{n-1} \frac{(-z)^k}{k!}.
\end{equation}
We have $H_n(z)\sim z^n$ at $z\to 0$ and $H_n(z)\sim z^{n-1}$ at $z\to\infty$, so the integral \eqref{DirectMellinDef} is convergent only for $-n < \Re s < -n+1$. Therefore it is the fundamental strip for $H_n(z)$. And all these functions $H_n(z)$ have the same Mellin image $\Gamma(s)$.

\begin{figure}
     \centering
     \begin{subfigure}[b]{0.34\textwidth}
         \centering
         \includegraphics[width=\textwidth]{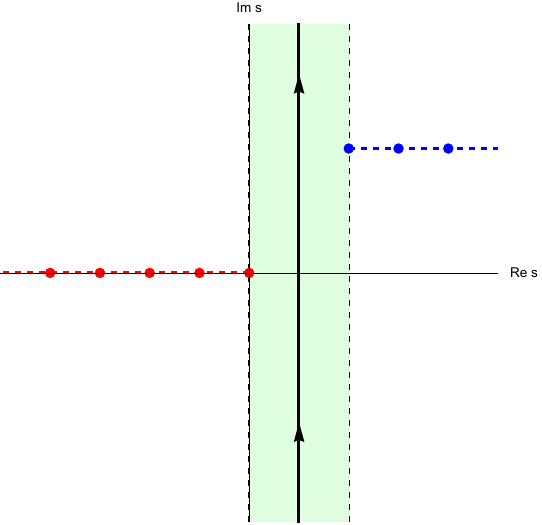}
         \caption{\footnotesize{Pole structure of \eqref{GamGamEq} for $\Re\a>0$: the MB contour is straight and is contained inside the fundamental strip (shown in green).}}
         \label{fig:ancont_1}
     \end{subfigure}
     \hfill
     \begin{subfigure}[b]{0.34\textwidth}
         \centering
         \includegraphics[width=\textwidth]{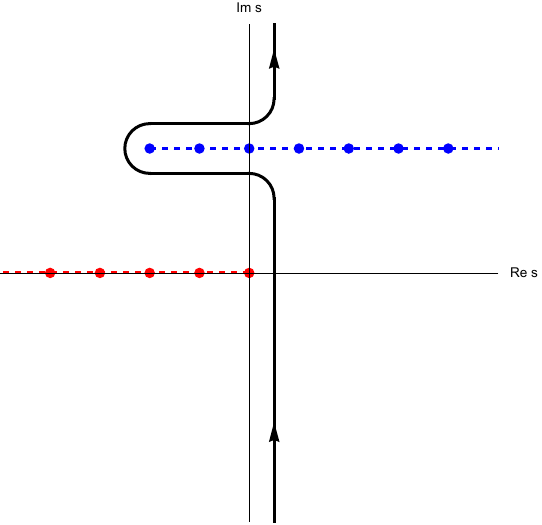}
         \caption{\footnotesize{Pole structure of \eqref{GamGamEq} for $\Re\alpha<0$ (and $\alpha\notin\bbZ_{\leq0}$): the fundamental strip is empty and the MB contour is curved.}}
         \label{fig:ancont_2}
     \end{subfigure}
     \hfill
     \begin{subfigure}[b]{0.34\textwidth}
         \centering
         \includegraphics[width=\textwidth]{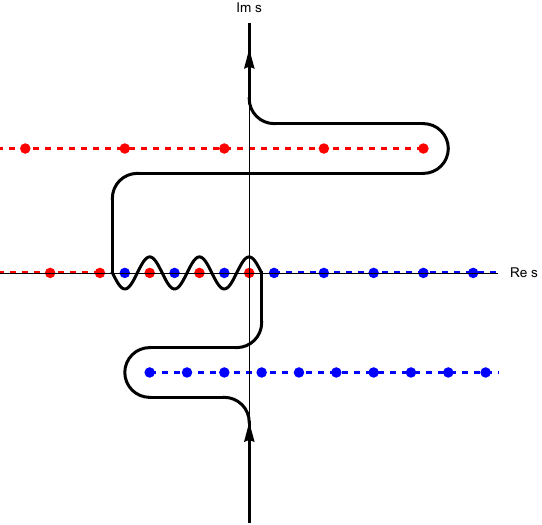}
         \caption{\footnotesize{MB contour in more complicated case of the type~\eqref{FoxImage}.}}
         \label{fig:ancont_3}
     \end{subfigure}
    \caption{\footnotesize{Analytic continuation of MB integral in parameter $\alpha$ and resulting non-splitting contour}}
        \label{fig:ancont}
\end{figure}

Secondly, for some functions $H(z)$ the direct Mellin transform may be not defined at all in the sense that the integral \eqref{DirectMellinDef} diverges for all values of the parameter $s$, and the fundamental strip of the function $H(z)$ is empty. However, even in this case the function $H(z)$ can be represented as an integral of the form \eqref{InverseMellinDef}, the integration in which will now be not over a straight vertical contour, but over its deformation that bypasses the singularities of the function $h(s)$. Moreover, if the original function $H_\alpha(z)$ depends on some external parameter $\alpha$, and in some range of $\alpha$ the integral \eqref{DirectMellinDef} diverges everywhere, then its ``Mellin image'' will still be given by the same function $h_\alpha(s)$, continued from the region of $\alpha$ in which the Mellin transform is well defined.

Another simple illustrative example: consider the function
\begin{equation}
H_\alpha(z) = \frac{\Gamma(\alpha)}{(1+z)^\alpha},
\end{equation}
where $\alpha\notin\bbZ_{\le0}$. It is easy to verify that for $\Re\alpha>0$ this function has a fundamental strip $0<\Re s<\Re\alpha$, inside which the integral \eqref{DirectMellinDef} converges and generates the following Mellin image
\begin{equation} \label{GamGamEq}
h_\alpha(s) = \int\limits_0^\infty dz\, \frac{\Gamma(\alpha) z^{s-1}}{(1+z)^\alpha} = \Gamma(s)\Gamma(\alpha-s).
\end{equation}
This function can be analytically continued to the entire complex plane except for the points $s=-n$ and $s=\alpha+n$, $n\in\bbZ_{\ge0}$, where the gamma functions have poles. Then the original function can be reconstructed by the formula \eqref{InverseMellinDef} with a straight contour lying in the fundamental strip (see Fig.~\ref{fig:ancont_1}).

However, for $\Re\alpha<0$ the integral \eqref{DirectMellinDef} diverges for any values of $s$---the fundamental strip is empty, and the direct Mellin transform is not defined. Nevertheless, even in this case the original function can be represented by an integral of the form \eqref{InverseMellinDef}
\begin{equation} \label{MBexample}
H_\alpha(z) = \int\limits_C \frac{ds\, z^{-s}}{2\pi i} \Gamma(s)\Gamma(\alpha-s),
\end{equation}
where the integration contour $C$ is now not straight, but curved and separating the poles of the gamma functions $\Gamma(s)$ and $\Gamma(\alpha-s)$ (see Fig.~\ref{fig:ancont_2}).

A completely analogous situation occurs in the much more general case of the well-known Fox $H$-functions $H_{p,q}^{m,n}\!\left[z \left|\,\begin{smallmatrix}(a, A) \\ (b, B) \end{smallmatrix} \right. \right]$, for which the Mellin image has the form
\begin{equation} \label{FoxImage}
h(s) = \frac{\prod\limits_{i=1}^m \Gamma(b_i + B_i s) \prod\limits_{j=1}^n \Gamma(1-a_j - A_j s)}{\prod\limits_{i=m+1}^q \Gamma(1-b_i - B_i s) \prod\limits_{j=n+1}^p \Gamma(a_j + A_j s)},
\end{equation}
where $A_j$ and $B_i$ are positive, and the corresponding arrangement of poles is shown schematically in Figure~\ref{fig:ancont_3}. If the leftward- and rightward-running poles do not coalesce, it is always possible to draw a contour $C$ that would separate them. Clearly, the parameters $a_j$ and $b_i$ may be such that this contour cannot be straightened. In this case, the integral \eqref{DirectMellinDef} will diverge for any value of $s$. However, by continuously varying the parameters $a_j$ and $b_i$, we can always pull the poles apart so that a non-empty fundamental strip appears. And in this strip, the integral \eqref{DirectMellinDef} will reproduce the expression \eqref{FoxImage}.

This observation is a rigorous justification of the procedure we use for regularizing divergent integrals by analytic continuation with respect to a parameter.

\subsection{Multiple MB integrals} \label{DefSubSec}

$N$-fold MB integrals define a class special functions of hypergeometric type of $N$ complex variables. Suppose that in addition to $N$, we are also given another integer $m$. We will assume that the index $a$ runs from 1 to $N$, and the index $i$ runs from 1 to $m$. Suppose we are given data of the three following types:
\begin{itemize}
\item vector $\bm{\alpha} = (\alpha_i)$ of length $m$ with nonzero integer components;

\item $(N\times m)$-matrix $A = (A_i^a)$ with real components, it can be decomposed into $m$ column vectors $\bm{A}_i = (A_i^a)$, which must be nonzero $\bm{A}_i \ne 0$;

\item finally, a vector $\bm{a} = (a_i)$ of length $m$ with arbitrary complex components.
\end{itemize}

First, we define a function that is the product of $m$ gamma factors with arguments linear in $N$ complex variables $\bm{s} = (s_a) \in\bbC^N$:
\begin{equation} \label{GammaProduct}
h\left(\bm{\alpha}, A, \bm{a} | \bm{s} \right) = \prod\limits_{i=1}^m \Gamma_i, \qquad \Gamma_i = \Gamma^{\alpha_i}(\bm{s} \cdot \bm{A}_i  - a_i),
\end{equation}
where $\bm{s} \cdot \bm{A}_i = s_a A_i^a$. Then a $N$-fold MB integral will look like
\begin{equation} \label{NfoldIntDef}
H\left(\bm{\alpha}, A, \bm{a} | \bm{z} \right) =  \int\limits_C \frac{d^Ns}{(2\pi i)^N}\, h(\bm{\alpha}, A, \bm{a} | \bm{s})\, \bm{z}^{-\bm s},
\end{equation}
where $\bm{z} = (z_a) \in\bbC^N$ and $\bm{z}^{-\bm s} = \exp(-\bm{s}\cdot\ln\bm{z})$ for $|\arg z_a| < \pi$.\footnote{Here and below for functions such as $\ln$, $|\ldots|$, $\arg$, etc. we agree to assume that $f(\bm{z}) = \big(f(z_1), \ldots, f(z_N)\big)$.} The integration here is over some $N$-cycle $C$, which will be characterized in detail below.

First of all, we note that the Mellin image $h(\bm{\alpha}, A, \bm{a} | \bm{z})$ \eqref{GammaProduct} does not depend on the order of the gamma factors $\Gamma_i = \Gamma^{\alpha_i}(\bm{s}\cdot\bm{A}_i - a_i)$. This means that the MB integral $H(\bm{\alpha}, A, \bm{a} | \bm{z})$ does not change under a simultaneous permutation of the components of the vectors $\bm{\alpha}$, $\bm{a}$ and the columns of the matrix $A$ (just as its value does not change under a simultaneous permutation of the arguments $\bm{z}$ and the rows of the matrix $A$). Since the role of the gamma factor $\Gamma_i$ is determined by the sign of the parameter $\alpha_i$, it is convenient to agree they are ordered such that the first $m_+$ factors are with positive powers $\alpha_i>0$, followed by $m_-$ factors with negative powers $\alpha_i<0$, $m_+ + m_- = m$. We will also denote $\calA_+ = \{\bm{A}_i \mid \alpha_i>0\}$.

The singularities of the function $h(\bm{s})$ \eqref{GammaProduct} are determined by the singularities of the $m_+$ gamma factors $\Gamma_i$ for which $\alpha_i>0$. Each of these factors $\Gamma_i$ generates an infinite family of polar hyperplanes of complex dimension $(N-1)$
\begin{equation} \label{Hyperplanes}
L_{i,n} = \{ \bm{s}\in\bbC^N \mid \bm{s} \cdot \bm{A}_i  = a_i - n \}.
\end{equation}

The $N$-cycle $C$ over which the integration~\eqref{NfoldIntDef} is performed is defined as follows: one can always choose the parameters $\bm{a}$ such that a straight vertical contour $\bm{\gamma} + i\bbR^N$ can be drawn such that for each $i = 1, \ldots, m_+$ it lies to the same side of all polar hyperplanes $L_{i,n}$ \eqref{Hyperplanes} (i.e. $\bm{\gamma}\cdot \bm{A}_i > \Re a_i$), then we choose it as $C$. If the parameters $\bm{a}$ become such that a straight vertical contour with the specified property ceases to exist, $C$ should be understood as a continuous deformation of $\bm{\gamma} + i\bbR^N$ that does not intersect the polar hyperplanes $L_{i,n}$.

The multiple MB integral \eqref{NfoldIntDef} can converge or diverge depending on the external parameter $\bm{z}$. The situation is governed by the following key invariant:
\begin{equation} \label{VarpiDef}
\varpi = \min_{\|\bm{x}\| = 1} \sum\limits_{i=1}^m \alpha_i\, |\bm{x} \cdot \bm{A}_i|,
\end{equation}
where $\bm{x}\in\bbR^N$. Using the Stirling formula \eqref{Stirling}, it is easy to show \cite{Zhdanov1998} that for $\varpi\le 0$ the integral \eqref{NfoldIntDef} diverges for any $\bm{z}$. If $\varpi >0$, it converges absolutely in the domain
\begin{equation}
\begin{split}
    U_\varpi = \{\bm{z}\in\bbC^N &\mid z_a\ne 0,\;\\ &|\arg z_a|<\pi,\; \|\arg\bm{z}\| < \tfrac{\pi}{2}\varpi \}
\end{split}
\label{Udomain}
\end{equation}
and defines an analytic function in it, which can then be analytically extended from $U_\varpi$ to all $\bbC^N$ except a certain singular surface of lower dimension (for Feynman integrals, it is called a ``Landau surface'').

For $\varpi>0$ and for the $N$-fold MB integral \eqref{NfoldIntDef} to be well-defined, it is necessary, in particular, that the number of gamma factors with positive powers be no less than the dimension $m_+ \ge N$, and that $\calA_+$ contain at least one set of $N$ vectors that forms a basis for the $\bm{s}$-space. For brevity, we will call such sets \emph{clusters} and denote them by subsets $I\subset\{1, \ldots, m_+\}$.

Finally, we note that the linear structure of the arguments of the gamma factors $\Gamma_i$ in \eqref{GammaProduct} is preserved under general affine transformations of $\bm{s}$-space. That is, under shifts by an arbitrary complex vector $\bm{l} = (l_a) \in \bbC^N$, $\bm{s} \mapsto \bm{s} + \bm{l}$, and under arbitrary invertible linear transformations defined by a real matrix $M = (M_b^a)$, $\bm{s} \mapsto \bm{s} M = s_b M_a^b$. This leads to the relations:
\begin{align}
H(\alpha_i, \bm{A}_i, a_i|\, \bm{z}) &=  \bm{z}^{\bm l}\; H(\alpha_i, \bm{A}_i, a_i + \bm{l}\cdot \bm{A}_i |\, \bm{z}) \label{Transf1} \\
&= \det M \; H(\alpha_i, M \bm{A}_i, a_i|\, \bm{u}), \label{Transf2}
\end{align}
where $M\bm{A}_i = M_b^a A_i^b$ and $\ln\bm{u} = M\ln\bm{z}$.

\subsection{Series representations} \label{SeriesSubSec}

\paragraph{Non-resonant and resonant cases.}

We will assume that the coefficients of the matrix $A$ introduced above (and so its column vectors $\bm{A}_i$) are fixed, while the complex parameters $\bm{a} = (a_i)$ are variable. For each cluster $I$, we denote $\bm{a}_I = (a_i)_{i\in I}$. Analogously, by $A_I$ we denote the square $(N\times N)$-matrix that results if we retain in $A$ only the columns $\bm{A}_i$ for which $i\in I$. Since they are linearly independent by definition, $\Delta_I = \det A_I \ne 0$. But then the $N$ polar hyperplanes of $L_{i,n}$ corresponding to a given cluster $I$ will intersect at an infinite family of points, which we will call \emph{simple poles},
\begin{equation} \label{ClusterPoints}
\bm{s}_I(\bm{n}) = (\bm{a}_I - \bm{n}) A_I^{-1},
\end{equation}
where $A_I^{-1}$ is the matrix inverse to $A_I$, and $\bm{n} \in \bbZ_{\ge0}^N$.

The following fundamental alternative holds:
\begin{itemize}
\item In the generic \emph{non-resonant} case (i.e., for almost all values of the parameter $\bm{a}$), the simple poles do not coincide with each other. In other words, there are no points at which more than $N$ polar hyperplanes $L_{i,n}$ intersect.

\item However, for some exceptional values of the parameter $\bm{a}$, several simple poles can coalesce, generating a point at which more than $N$ polar hyperplanes $L_{i,n}$ intersect. Such an exceptional case is called \emph{resonant}.
\end{itemize}

Although the resonant case is exceptional, it is extremely important from a physical point of view. For example, dimensionally and analytically regularized Feynman integrals belong to the non-resonant case, but removing the regularization leads to the resonant case. On the other hand, the generic non-resonant case is much simpler, since then a multidimensional residue at a simple pole can be understood simply as repeatedly taking $N$ ordinary 1-dimensional residues with respect to the variables $s_a$, and if the simple poles coalesce, this requires the use of the local Grothendieck residue. Moreover, the non-resonant case is in a sense more fundamental, since the resonant case can be always obtained from it by taking the appropriate limit with respect to the parameter $\bm{a}$. Therefore, it is correct to first consider the generic non-resonant case and only then move on to the resonant one.

\paragraph{Cluster series.}

The study of the non-resonant case leads to a very important concept of a \emph{cluster series}. We will call this the formal series obtained by summing the residues of the integrand in \eqref{NfoldIntDef} at the simple poles $\bm{s}_I(\bm{n})$ \eqref{ClusterPoints} generated by some given cluster $I$:
\begin{equation} \label{ClusterResidues}
\calH_I(\bm{z}) = \sum\limits_{\bm{n}\in\bbZ_{\ge0}^N} \Res\limits_{\bm{s} = \bm{s}_I(\bm{n})} \left(h(\bm{s})\, \bm{z}^{-\bm s}\right).
\end{equation}

In the simplest case when $\alpha_i = 1$ for all $i\in I$, such cluster series reduces to the Horn series:
\begin{equation} \label{Horn1}
\calH_I(\bm{z}) = \Delta_I^{-1} \bm{u}^{-\bm{a}_I} \sum\limits_{\bm{n}\in\bbZ_{\ge0}^N} c_{\bm n} \frac{(-\bm{u})^{\bm n}}{\bm{n}!},
\end{equation}
where
\begin{align}
&\ln\bm{u} = A_I^{-1} \ln\bm{z}, \quad \ln\bm{z} = A_I \ln\bm{u}, \label{TransfOfVariables} \\
&c_{\bm n} = \prod\limits_{i\notin I} \Gamma^{\alpha_i}(-\bm{n}\cdot \bm{B}_i - b_i), \label{Horn2} \\
&\bm{B}_i = A_I^{-1}\bm{A}_i, \quad b_i = a_i - \bm{a}_I A_I^{-1} \bm{A}_i.
\end{align}

In the case when $\alpha_i = 2, 3, \ldots$ for some $i\in I$, the corresponding factors $\Gamma_i$ have poles of higher orders. Then we will have a generalized logarithmic series (containing, along with the powers $\bm{u}^{\bm n}$, also the powers of the logarithms $\ln u_a$) with polygamma functions $\psi^{(k)}(n_i)$ in its coefficients. Although deriving the corresponding expressions does not present any difficulty, they do not appear in this work and are rather cumbersome, so we will not present them here.

\paragraph{Reduction of a 2-fold MB integral to combination of cluster series.}

Using a multivariate generalization of the residue theorem, an $N$-fold MB integral $H(\bm{\alpha}, A, \bm{a} |\, \bm{z})$ \eqref{NfoldIntDef} can be (generally speaking, ambiguously) represented as combinations of the above-introduced cluster series $\calH_I(\bm{z})$ \eqref{ClusterResidues}.

In the trivial case $N=1$, the MB integral is simply the well-known Fox $H$-function \eqref{FoxImage}, and there are only two representations: we can ``close the contour'' $C$ either to the right or to the left and reduce the integral to the sum of residues at all leftward- or rightward-running poles, respectively, resulting in expansions near the points $z=0$ or $z=\infty$.

In fact, the case $N=2$ is also almost trivial in the sense that these representations can also be constructed without any difficulty. The corresponding algorithm was basically built in \cite{Zhdanov1998}. However, already for $N=3$, the situation becomes significantly more complex, and some subtle points and pitfalls arise.

However, in the present paper, only a single 3-fold integral appears, and one that does not encounter the difficulties noted. Therefore, we will limit ourselves to briefly presenting the algorithm for $N=2$ without justification, and will not touch on the case $N>2$ at all. 

In the case $N=2$, each cluster is a pair of nonzero, non-collinear real vectors $\bm{A}_i$ and $\bm{A}_j$ on the plane $\bbR^2$, which we will denote by $\{ij\}$. The conical hull of these vectors is a sector with an angle less than $\pi$, which we will denote by $S_{ij}$. Among all sectors defined by different clusters of vectors from $\calA_+$, there exist minimal ones, i.e., ones such that no other vector from $\calA_+$ falls within them.

So, each such minimal sector $S$ defines its own representation of a 2-fold MB integral $H(\bm{\alpha}, A, \bm{a} |\, \bm{z})$ as the sum of cluster series $\calH_{ij}(\bm{z})$, over all clusters $\{ij\}$ such that the corresponding sectors $S_{ij}$ contain the given minimal sector $S$:
\begin{equation} \label{SumOverClusters}
H(\bm{\alpha_i}, A, \bm{a} |\, \bm{z}) \approx \sum\limits_{\{ij\}\colon S\subset S_{ij}} \calH_{ij}(\bm{z}).
\end{equation}

\paragraph{Convergence of series.}

In the formula \eqref{SumOverClusters} above, we put the sign $\approx$ instead of the equality sign, since until now we have not asked the question: do the cluster series $\calH_{ij}(\bm{z})$ on the right-hand side of \eqref{SumOverClusters} converge and, if so, for what $\bm{z}$? The general answer to this question is controlled by the vector parameter
\begin{equation} \label{Param1}
\bm{A} = \sum\limits_{k=1}^m \alpha_k \bm{A}_k,
\end{equation}
and for the case $N=2$ it is also very simple. The cluster series $\calH_{ij}(\bm{u})$ is absolutely convergent everywhere if the vector $\bm{A}$ lies inside the sector $S_{ij}$\footnote{It is important that the variable $\bm{u}$ \eqref{TransfOfVariables} is used here, not $\bm{z}$. Accordingly, the singularities with respect to the variables $z_a$ lie either at infinity or at zero.}, and asymptotic (divergent) if $\bm{A}$ lies outside of $S_{ij}$. If it lies exactly on the boundary of $S_{ij}$, then it converges in some region $U_{ij}$ and has a singularity on its boundary.

Accordingly, only three cases are possible:
\begin{itemize}
\item \emph{Unbalanced} case: the vector $\bm{A} \ne0$ lies in the minimal sector $S_\mathrm{conv}$. Then all cluster series in the \eqref{SumOverClusters} representation for this sector $S_\mathrm{conv}$ will be absolutely convergent, and for any other minimal sector $S$, some of the cluster series in the corresponding \eqref{SumOverClusters} representation will be asymptotic.

\item \emph{Balanced} case: $\bm{A} = 0$. Then, for each minimal sector $S$, the \eqref{SumOverClusters} representation is well-defined in its domain, which is the intersection of the corresponding domains $U_{ij}$, and these representations analytically continue each other.

\item Finally, in the intermediate case, the vector $\bm{A} \ne0$ lies exactly on the boundary of the two minimal sectors $S_1$ and $S_2$. Then the corresponding representations \eqref{SumOverClusters} for these two sectors are each defined in their own domains $U_1$ and $U_2$ and analytically continue each other. And for any other sector $S$, the corresponding representation will contain asymptotic cluster series.
\end{itemize}

\paragraph{Pinches and the logarithmic case}

So far, we have considered the generic non-resonant case, where the parameters $\bm{a}$ are such that the simple poles $\bm{s}_I(\bm{n})$ \eqref{ClusterPoints} do not coalesce. Then, their residues form cluster series \eqref{ClusterResidues}, and different clusters do not interact with each other, leading to a sum over clusters \eqref{SumOverClusters}.

However, for some values of the parameter $\bm{a}$, some of the coefficients in the cluster series \eqref{ClusterResidues} may become infinite. This situation always corresponds to a coalescence of several simple poles $\bm{s}_I(\bm{n})$ \eqref{ClusterPoints} generated by different clusters $I$. Thus, the corresponding resonant case can be understood as an interaction between clusters. The following fundamental alternative is possible:
\begin{itemize}
\item It may happen that the $N$-cycle $C$ is pinched between coalescing simple poles. So this situation is called a \emph{``pinch''}. Then the divergence of the coefficients in the cluster series \eqref{ClusterResidues} leads to the entire MB integral $H(\bm{\alpha}, A, \bm{a} |\, \bm{z})$ becoming infinite, i.e. it has a pole as a function of the parameter $\bm{a}$.

\item If no pinch occurs, the coalescing simple poles form a higher order pole, the residue at which should be understood as a local Grothendieck residue. Its calculation leads to the appearance of logarithms, which is why this case is called \emph{logarithmic} in the literature. Although divergences arise in the coefficients of the formal cluster series \eqref{ClusterResidues}, they cancel each other out in such a way that the entire MB integral $H(\bm{\alpha}, A, \bm{a} |\, \bm{z})$ remains finite. This simply means that in the logarithmic case, clusters generating coalescing simple poles can no longer be considered independent.
\end{itemize}

\section{General structure of basis and complete kernels} \label{StructureSec}

In this section, we consider the structure of basis $\bbB_\alpha[f | \sigma]$ \eqref{defLf} and complete massive $\bbW_\alpha[f | \sigma, m^2]$ \eqref{BHtransform} kernels. While the general statements presented here clarify the specific examples discussed later in Sec.~\ref{NonresSec}, they are ``empirical'' in nature in that they simply generalize observations on analytic expressions in these specific examples. The question of their rigorous justification and applicability is certainly of considerable interest, but it remains outside the scope here.

First of all, note that it is also convenient to introduce a third family of functions
\begin{align}
&\bbM_\alpha[f | m^2] = \frakL_f M_\alpha(\tau, m^2), \label{bbMdef} \\
&M_\alpha(\tau, m^2) = \frac{\tau^{-\alpha}}{(4\pi)^{d/2}} \exp\left(- \tau m^2 \right).
\end{align}
We call them \emph{complementary kernels} because their properties are indeed complementary in many aspects to those of the basis kernels $\bbB_\alpha[f | \sigma]$. In particular, if the integral \eqref{defLf} always converges well at the UV limit $\tau=0$, but can diverge at the IR limit $\tau=\infty$ for sufficiently small $\Re\alpha$, then the situation with the integral \eqref{bbMdef} is exactly the opposite: it converges well at the IR limit $\tau=\infty$, but can begin to diverge on the UV limit $\tau=0$ for sufficiently large $\Re\alpha$ (and in this case, too, it requires regularization via analytic continuation).

Next, we apply our trick with term-by-term integration of asymptotic expansions to the definition of complete massive kernels $\bbW_\alpha[f | \sigma, m^2]$ \eqref{BHtransform}. If in the expression $W_\alpha(\tau, \sigma, m^2)$ \eqref{IRregKernel} we expand the exponential $\exp(-\tau m^2)$ into a power series corresponding to the UV limit $\tau\to0$, apply the integral transform $\frakL_f$ \eqref{frakLtransformEq} to it term-by-term, and use the definition \eqref{defLf}, then we immediately obtain the expansion
\begin{equation} \label{UVexpansion}
\bbW_\alpha[f | \sigma, m^2] \xrightarrow{\text{UV}} \sum\limits_{k=0}^\infty \frac{(-m^2)^k}{k!}\, \bbB_{\alpha-k}[f | \sigma].
\end{equation}
On the contrary, if in \eqref{IRregKernel} we expand the exponential $\exp(-\sigma/2\tau)$ into a power series corresponding to the IR limit $\tau\to\infty$, apply $\frakL_f$ \eqref{frakLtransformEq} to it term-by-term, and use the definition of \eqref{bbMdef}, we obtain a completely different expansion
\begin{equation} \label{IRexpansion}
\bbW_\alpha[f | \sigma, m^2] \xrightarrow{\text{IR}} \sum\limits_{n=0}^\infty \frac{(-\sigma/2)^n}{n!}\, \bbM_{\alpha+n}[f | m^2].
\end{equation}

Although these transforms involve manipulations with divergent series and integrals, as we explained in \cite{BKWletter, BKW25a}, the remarkable fact is that the expressions \eqref{UVexpansion}-\eqref{IRexpansion} are not meaningless at all, but rather somehow determine the structure of the complete massive kernel $\bbW_\alpha[f | \sigma, m^2]$ \eqref{BHtransform}. In the simplest case of complex power $f(\hat F) = \hat F^{-\mu}$, as we discussed in detail in \cite{BKWletter, BKW25a} and will briefly reproduce below in Subsec.~\ref{ComplexPowerSubsec}, the complete massive kernel $\bbW_\alpha[F^{-\mu} | \sigma, m^2]$ is exactly equal to the sum of the two series \eqref{BCdecomposition} corresponding to the expressions \eqref{UVexpansion}-\eqref{IRexpansion}. In the general case of an arbitrary operator function $f(\hat F)$, the situation is slightly more complicated.

By carefully comparing the series representations obtained for specific functions $f(\hat F)$ below in Sec.~\ref{NonresSec}, we can verify that the expression \eqref{BHtransform} is formed by series of three types: some of them are present simultaneously in both expansions \eqref{UVexpansion}-\eqref{IRexpansion}---we will call them \emph{regular}; while others---we will call them \emph{singular}---are present in only one of them. These, in turn, are clearly divided into those coming from the UV region, i.e., appearing only in the expansion \eqref{UVexpansion}, or from the IR region, i.e., appearing only in the expansion \eqref{IRexpansion}. We will denote the corresponding contributions generated by terms of these three types by the superscripts `reg', `UV', and `IR'.

It is also convenient,  in basis $\bbB_\alpha[f | \sigma]$ \eqref{defLf} and complementary $\bbM_\alpha[f | m^2]$ \eqref{bbMdef} kernels, to separate the contributions that generate regular and singular terms in complete kernels $\bbW_\alpha[f | \sigma, m^2]$ \eqref{BHtransform}. By definition, the basis kernel $\bbB_\alpha[f | \sigma]$ will contain only regular and UV singular contributions, while the complementary kernel $\bbM_\alpha[f | m^2]$, conversely, will contain only regular and IR singular contributions. As a result, we obtain the following decompositions:
\begin{align}
\bbB_\alpha[f | \sigma] &= \bbB_\alpha^\text{reg} + \bbB_\alpha^\text{UV}, \\
\bbM_\alpha[f | m^2] &= \bbM_\alpha^\text{reg} + \bbM_\alpha^\text{IR}, \\
\bbW_\alpha[f | \sigma, m^2] &= \bbW_\alpha^\text{reg} + \bbW_\alpha^\text{UV} + \bbW_\alpha^\text{IR}.
\end{align}
In this case, the corresponding singular parts will be related by the relations
\begin{align}
\bbW_\alpha^\text{UV}[f | \sigma, m^2] &= \sum\limits_{k=0}^\infty \frac{(-m^2)^k}{k!} \bbB_{\alpha-k}^\text{UV}[f | \sigma], \label{FundExpansion1} \\
\bbW_\alpha^\text{IR}[f | \sigma, m^2] &= \sum\limits_{n=0}^\infty \frac{(-\sigma/2)^n}{n!} \bbM_{\alpha+n}^\text{IR}[f | m^2], \label{FundExpansion2}
\end{align}
and for the regular part of complete massive kernels, two types of relations will hold:
\begin{align}
\bbW_\alpha^\text{reg}[f | \sigma, m^2] &= \sum\limits_{k=0}^\infty \frac{(-m^2)^k}{k!} \bbB_{\alpha-k}^\text{reg}[f | \sigma], \nonumber \\
&= \sum\limits_{n=0}^\infty \frac{(-\sigma/2)^n}{n!} \bbM_{\alpha+n}^\text{reg}[f | m^2]. \label{FundExpansion3}
\end{align}

It naturally arises a hypothesis that the compatibility of these relations must be ensured by the existence of some fourth family of objects $\bbL_\alpha[f]$ such that
\begin{align}
\bbM_\alpha^\text{reg}[f | m^2] &= \sum\limits_{k=0}^\infty \frac{(-m^2)^k}{k!} \bbL_{\alpha-k}[f], \label{FundExpansion4} \\
\bbB_\alpha^\text{reg}[f | \sigma] &= \sum\limits_{n=0}^\infty \frac{(-\sigma/2)^n}{n!} \bbL_{\alpha+n}[f], \label{FundExpansion5}
\end{align}
where $\bbL_\alpha[f]$ does not depend neither on $\sigma$ nor $m^2$. We will call these objects the \emph{regular kernels}. The specific examples discussed further in Sec.~\ref{NonresSec} confirm this assumption.

To summarize this discussion, we have learned how to reduce the ``two-variable'' function $\bbW_\alpha[f | \sigma, m^2]$ to two ``one-variable'' functions $\bbB_\alpha^\text{UV}[f | \sigma]$ and $\bbM_\alpha^\text{IR}[f | m^2]$ and one ``constant'' $\bbL_\alpha[f]$, which is a significant simplification.

The expansions \eqref{FundExpansion3}-\eqref{FundExpansion5} imply the following structure of limits for regular contributions:
\begin{equation} \label{FundLimits}
\xymatrix{& \bbW_\alpha^\text{reg}[f | \sigma, m^2] \ar@{->}[dl]^{m^2\to0} \ar@{->}[dr]_{\sigma\to0} & \\
\bbB_\alpha^\text{reg}[f | \sigma] \ar@{->}[dr]_{\sigma\to0} && \bbM_\alpha^\text{reg}[f | m^2] \ar@{->}[dl]^{m^2\to0}. \\
& \bbL_\alpha[f] &
}\end{equation}

At the same time, the coincidence limit $\sigma\to0$ for the ultraviolet part of the basis kernel $\bbB_\alpha^\text{UV}[f | \sigma]$ behaves simply as a power of $\sigma$. And similarly, the massless limit $m^2\to0$ for the infrared part of the complementary kernel $\bbM_\alpha^\text{IR}[f | m^2]$ behaves simply as a power of $m^2$. That is, they either tend to infinity or vanish depending on the values of the parameter $\alpha$. It is for this reason that we call the different contributions ``regular'' and ``singular''.

\section{Examples in non-resonant case} \label{NonresSec}

In this section, we will consider the functions obtained in \cite{BKW25a} for the non-resonant case (the resonant case will be discussed further in Sec.~\ref{ResonantSec}). First, we briefly reproduce in Subsec.~\ref{ComplexPowerSubsec} the results for the complex power $\hat F^{-\mu}$ discussed in detail earlier in \cite{BKWletter, BKW25a}. We then consider successively the hybrid function $\exp(-\tau\hat F^\nu)/\hat F^\mu$ in Subsec.~\ref{HybridSubsec}, the resolvent $1/(\hat F^\mu + \lambda)$ in Subsec.~\ref{ResolventSubsec}, and finally, the most complex of our operator functions, $\exp(-\tau\hat F^\nu)/(\hat F^\mu + \lambda)$, in Subsec.~\ref{TheMostDifficultSubsec}.

In accordance with our general scheme outlined in the previous Sec.~\ref{StructureSec}, we will interpret the series arising in these examples as representing either regular, or UV and IR singular contributions (and denote this by the corresponding superscripts `reg', `UV', and `IR'). In each case, it is easy to verify that these contributions indeed satisfy our fundamental expansions \eqref{FundExpansion1}-\eqref{FundExpansion5}. Therefore, we will not justify our interpretation in detail---in each case, it is based on the indicated easily verifiable correspondence. In Subsec.~\ref{TheMostDifficultSubsec}, we will slightly extend this interpretation, further distinguishing the various contributions among the regular terms.

\subsection{The complex power $\hat F^{-\mu}$} \label{ComplexPowerSubsec}

First of all, let us recall that the basis and complementary kernels for the operator power $\hat F^{-\mu}$ are not represented by MB integrals, but are just powers of $\sigma$ and $m^2$:
\begin{align}
\bbB_\alpha\!\big[F^{-\mu} \big| \sigma\big] &= \frac{\Gamma\left(\alpha-\mu\right)}{(4\pi)^{d/2}\Gamma(\mu)} \left(\frac{\sigma}{2}\right)^{\mu-\alpha}, \label{CompPowFunctionsApp} \\
\bbM_\alpha\!\big[F^{-\mu} \big| m^2\big] &= \frac{\Gamma\left(\mu-\alpha\right)}{(4\pi)^{d/2}\Gamma(\mu)}\, m^{2(\alpha-\mu)}. \label{CompPowFunctions2App}
\end{align}
(It is convenient to think of them as trivial MB integrals ``of $N=0$ variables'').

Further, the complete massive kernel for the power $\hat F^{-\mu}$ is expressed through the Bessel-Clifford function of the second kind:
\begin{align}
&\bbW_\alpha\!\big[F^{-\mu} \big| \sigma, m^2\big] = \frac{2m^{2(\alpha-\mu)}}{(4\pi)^{d/2} \Gamma(\mu)}\; \calK_{\alpha-\mu}(z), \label{GreenIRregApp} \\
&\calK_\alpha(z) = \frac{1}{2} \int\limits_0^\infty dt\, t^{-\alpha-1}\, \exp\left(-t - \frac{z}{t}\right), \quad z = \frac{\sigma m^2}{2}. \label{BCdef}
\end{align}

As we explained in detail in \cite{BKWletter, BKW25a}, the function $\calK_\alpha(z)$ (and hence the complete kernel $\bbW_\alpha\!\big[F^{-\mu} \big| \sigma, m^2\big]$) can be represented as a single MB integral:
\begin{align}
&\calK_\alpha(z) = \int\limits_C \frac{ds}{2\pi i}\, z^{-s} \kappa_\alpha(s), \label{BCintegral} \\
&\kappa_\alpha(s) = \frac{1}{2}\Gamma(s)\Gamma(s-\alpha). \label{BCMeelinBarnes2}
\end{align}
Further, we will systematically use this notation throughout: we will use a capital base symbol for the MB integral (e.g. $\calK$), and a lowercase base symbol for its Mellin image (e.g. $\kappa$), which is a product of gamma factors. We will not write out the MB integrals of type \eqref{BCintegral} themselves, implicitly assuming throughout that the corresponding pairs of functions are related by the $N$-fold inverse Mellin transform \eqref{NfoldIntDef}.

The Mellin image $\kappa_\alpha(s)$ has two rows of leftward-running poles $s_k = -k$ and $s_k = \alpha-k$, $k\in\bbZ_{\ge0}$. In the terminology of the general theory presented in Subsec.~\ref{SeriesSubSec}, these rows represent precisely the two clusters. Accordingly, the non-resonant case considered here means that these poles do not coalesce, i.e., that $\alpha\notin\bbZ$. Then we have a unique representation of the function $\calK_\alpha(z)$ as a series for $z\to0$, given by the sum of two cluster series. Importantly, these series can also be obtained in another way, namely, by integrating into \eqref{BCdef} the asymptotics of the integrand as $t\to0$ and $t\to\infty$. Accordingly, we interpret these cluster series as the UV and IR parts of the total expansion of the function, which we denote by superscripts. We have:
\begin{align}
\calK_\alpha(z) &= \calK_\alpha^\text{IR}(z) + \calK_\alpha^\text{UV}(z), \label{BCdecomposition} \\
\calK_\alpha^\text{UV}(z) &= \frac{1}{2} \sum\limits_{k=0}^\infty \frac{(-1)^k}{k!} \int\limits_0^\infty dt\; t^{k-\alpha-1} e^{-z/t} \nonumber \\
&= \frac{z^{-\alpha}}{2} \sum\limits_{k=0}^\infty \Gamma(\alpha-k) \frac{(-z)^k}{k!},  \label{BCseries2} \\
\calK_\alpha^\text{IR}(z) &= \frac{1}{2} \sum\limits_{k=0}^\infty \frac{(-z)^k}{k!} \int\limits_0^\infty dt\; t^{-\alpha-k-1} e^{-t} \nonumber \\
&= \frac{1}{2} \sum\limits_{k=0}^\infty \Gamma(-\alpha-k) \frac{(-z)^k}{k!}. \label{BCseries1}
\end{align}
In the parameter region $\Re(\alpha-\mu)>0$, which we call \emph{relevant}, the UV component \eqref{BCseries2} dominates in \eqref{BCdecomposition}, while in the parameter range $\Re(\alpha-\mu)<0$, which we call \emph{irrelevant}, the IR component \eqref{BCseries1} dominates.

Substituting these expressions into the formula \eqref{GreenIRregApp} and comparing the answers with \eqref{CompPowFunctionsApp}-\eqref{CompPowFunctions2App}, we can easily verify the validity of the fundamental expansions \eqref{FundExpansion1}-\eqref{FundExpansion2}. This means that for the complex power $\hat F^{-\mu}$, the regular part \eqref{FundExpansion3} is absent
\begin{equation}
\bbL_\alpha[F^{-\mu}] = 0,
\end{equation}
and the complete massive kernel $\bbW_\alpha[ F^{-\mu} | \sigma, m^2]$ is simply the sum of its UV and IR singular parts. As we will see below, this situation is, in a certain sense, atypical.

\subsection{The hybrid function $\exp(-\tau\hat F^\nu)/\hat F^\mu$} \label{HybridSubsec}

In this subsection, we consider the operator function $\exp(-\tau\hat F^\nu)/\hat F^\mu$, which we call ``hybrid'' in the sense that it interpolates between the operator exponential $\exp(-\tau\hat F^\nu)$ and the complex power $\hat F^{-\mu}$. For this function, the two cases are physically very different, depending on the sign of the parameter $\nu$: if $\nu>0$, our function behaves well in the UV region and poorly in the IR region, and for $\nu<0$, the situation is the opposite. We will consider both of these cases, which will allow us to obtain some physical interpretation of the UV and IR singular contributions.

\paragraph{The basis and complementary kernels.}

They are of the form (see \cite{BKW25a})\footnote{To avoid confusion, we note that since we initially aimed to make $\calE_{\nu,\alpha}^{(0)}(z)$ as similar as possible to a usual exponential when defining these functions, $\varepsilon_{\nu,\alpha}^{(\mu)}(s)$ is the Mellin image not of $\calE_{\nu,\alpha}^{(\mu)}(z)$, but of $\calE_{\nu,\alpha}^{(\mu)}(-z)$.}:
\begin{align}
&\bbB_\alpha\!\Big[ \frac{e^{-\tau F^\nu}}{F^\mu} \Big| \sigma\Big] = \frac{\tau^\frac{\mu-\alpha}{\nu}}{(4\pi)^{d/2}}\; \calE_{\nu,\alpha}^{(\mu)}(-z), \label{bbK_mn_from_bbK_0n} \\
&\varepsilon_{\nu,\alpha}^{(\mu)}(s) = \frac{\Gamma(s)\, \Gamma\!\left(\frac{\alpha-\mu-s}{\nu}\right)}{\nu\Gamma(\alpha-s)}, \qquad z = \frac{\sigma}{2\tau^{1/\nu}}, \label{calKMellin}
\end{align}
and
\begin{align}
&\bbM_\alpha\!\Big[ \frac{e^{-\tau F^\nu}}{F^\mu} \Big| m^2\Big] = \frac{\tau^\frac{\mu-\alpha}{\nu}}{(4\pi)^{d/2}}\; \tilde\calE_{\nu,\alpha}^{(\mu)}(z), \label{bbMtildeE} \\
&\tilde\varepsilon_{\nu,\alpha}^{(\mu)}(s) = \frac{\Gamma(s)\, \Gamma\!\left(\frac{s+\alpha-\mu}{\nu}\right)}{\nu\Gamma(s+\alpha)}, \qquad z = m^2\tau^{1/\nu}. \label{calKMellin2}
\end{align}

The image $\varepsilon_{\nu,\alpha}^{(\mu)}(s)$ \eqref{calKMellin} has two rows of poles: $s_k = -k$ and $s_k = \alpha - \mu + \nu k$, $k\in\bbZ_{\ge0}$. These two rows, in the terminology of Subsec.~\ref{SeriesSubSec}, represent two clusters and generate two corresponding cluster series:
\begin{equation} \label{SimpleSeries} \begin{aligned}
&\calH_1^\text{reg}(z) = \sum\limits_{k=0}^\infty \frac{\Gamma\!\left(\frac{k+\alpha-\mu}{\nu}\right)}{\nu\Gamma(k+\alpha)} \frac{(-z)^k}{k!}, \\
&\calH_2(z) = -z^{\mu-\alpha} \sum\limits_{k=0}^\infty \frac{\Gamma(\nu k+\alpha-\mu)}{\Gamma(\mu-\nu k)} \frac{(-z^{-\nu})^k}{k!}.
\end{aligned} \end{equation}
In this formula, we have not yet placed any superscript on the series $\calH_2(z)$, since its interpretation, as we will now see, depends on the sign of the parameter $\nu$.

If $\nu>0$, one row is leftward-running, and the other row is rightward-running. Then, the closure of the integration contour to the right and left generates two expansions of the function $\calE_{\nu,\alpha}^{(\mu)}(-z)$ as $z\to0$ and $z\to\infty$, and both cluster series \eqref{SimpleSeries} must be interpreted as two different representations of the regular contribution, which we will denote by overline:
\begin{equation} \label{CalEmu2}
\calE_{\nu,\alpha}^{(\mu)}(-z) = \calH_1^\text{reg}(z) = \calH_2^{\overline{\text{reg}}}(z).
\end{equation}
It is easy to show \cite{Wach2} that for $\nu>1/2$ the first series converges everywhere, while the second is asymptotic. For $0<\nu<1/2$, the situation is reversed. The case $\nu=1/2$ is balanced: the function has a singularity at $z=1/4$, therefore, the first series is convergent inside the circle $|z|<1/4$, while the second series is convergent outside of it. (It is also important that, for certain parameters $\nu$ and $\mu$, some poles of $\Gamma((\alpha-\mu-s)/\nu)$ in the numerator of \eqref{calKMellin} can cancel out with the poles of $\Gamma(\alpha-s)$ in the denominator.)

If $\nu<0$, both rows of poles are leftward-running. Therefore, there is a unique representation of the function $\calE_{\nu,\alpha}^{(\mu)}(-z)$ as the sum of two (everywhere converging) cluster series \eqref{SimpleSeries}, the first of which should be interpreted as the regular contribution, and the second as the UV singular contribution:
\begin{equation} \label{CalEmu2prime}
\calE_{\nu,\alpha}^{(\mu)}(-z) = \calH_1^\text{reg}(z) + \calH_2^\text{UV}(z).
\end{equation}
For $z\to0$, in the irrelevant region $\Re(\alpha-\mu)<0$, the leading term will come from the first series, and in the relevant region $\Re(\alpha-\mu)>0$, from the second one.

For the complementary kernel \eqref{bbMtildeE} the situation looks dual. Namely, the image $\tilde\varepsilon_{\nu,\alpha}^{(\mu)}(s)$ \eqref{calKMellin2} also has two rows of poles: $s_k = -k$ and $s_k = \mu - \alpha - \nu k$, which also generate two cluster series
\begin{equation} \label{tildeSimpleSeries} \begin{aligned}
&\tilde\calH_1^\text{reg}(z) = \sum\limits_{k=0}^\infty \frac{\Gamma\!\left(\frac{\alpha-\mu-k}{\nu}\right)}{\nu\Gamma(\alpha-k)} \frac{(-z)^k}{k!}, \\
&\tilde\calH_2(z) = z^{\alpha-\mu} \sum\limits_{k=0}^\infty \frac{\Gamma(\mu-\alpha-\nu k)}{\Gamma(\mu-\nu k)} \frac{(-z^\nu)^k}{k!}.
\end{aligned} \end{equation}

For $\nu>0$, both rows are leftward-running. Then, for the function $\tilde\calE_{\nu, \alpha}^{(\mu)}(z)$, there is a unique representation as the sum of two (everywhere converging) cluster series \eqref{tildeSimpleSeries}, the first of which we interpret as the regular contribution, and the second as the IR singular contribution:
\begin{equation} \label{tildeEexp}
\tilde\calE_{\nu, \alpha}^{(\mu)}(z) = \tilde\calH_1^\text{reg}(z) + \tilde\calH_2^\text{IR}(z). 
\end{equation}

Conversely, for $\nu<0$, one of the rows is leftward-running, and the other is rightward-running. Therefore, the two cluster series now correspond to two expansions of the function $\tilde\calE_{\nu, \alpha}^{(\mu)}(z)$ as $z\to0$ and $z\to\infty$, and both of them represent the regular contribution:
\begin{equation} \label{tildeEexpPrime}
\tilde\calE_{\nu, \alpha}^{(\mu)}(z) = \tilde\calH_1^\text{reg}(z) = \tilde\calH_2^{\overline{\text{reg}}}(z).
\end{equation}
The series $\tilde\calH_2^{\overline{\text{reg}}}(z)$ converges everywhere except $z=0$, and the series $\tilde\calH_1^\text{reg}(z)$ is asymptotic.

Next, substituting the expansions \eqref{SimpleSeries} and \eqref{tildeSimpleSeries} into the formulas \eqref{bbK_mn_from_bbK_0n} and \eqref{bbMtildeE} and comparing the resulting answers, we easily find a regular kernel $\bbL_\alpha[f]$ such that two limits from \eqref{FundLimits} hold. The corresponding expression is:
\begin{equation} \label{NulLKernellAppEq1}
\bbL_\alpha\!\Big[ \frac{e^{-\tau F^\nu}}{F^\mu} \Big] = \frac{\Gamma\left(\frac{\alpha-\mu}{\nu}\right)}{(4\pi)^{d/2}\nu\Gamma(\alpha)} \tau^\frac{\mu-\alpha}{\nu}.
\end{equation}

Note that when calculating the massless limit $m^2\to0$ for $\bbM_\alpha^\text{reg}$, we use the series $\tilde\calH_1^\text{reg}(z)$ in the variable $z = m^2\tau^{1/\nu} \to0$. But this means that for $\nu>0$, the same expression also describes the limit $\tau\to0$:
\begin{equation}
\bbM_\alpha^\text{reg}\!\Big[ \frac{e^{-\tau F^\nu}}{F^\mu} \Big| m^2 \Big] \xrightarrow[m^2, \tau \to0]{} \bbL_\alpha\!\Big[ \frac{e^{-\tau F^\nu}}{F^\mu} \Big].
\end{equation}
Similarly, for the IR singular part $\bbM_\alpha^\text{IR}$, we use the series $\tilde\calH_2^\text{IR}(z)$ in the same variable, from which we immediately obtain the limit:
\begin{equation}
\bbM_\alpha^\text{IR}\!\Big[ \frac{e^{-\tau F^\nu}}{F^\mu} \Big| m^2 \Big] \xrightarrow[m^2, \tau \to0]{} \bbM_\alpha\!\big[ F^{-\mu} \big| m^2 \big].
\end{equation}
In these two formulas, the notation $m^2, \tau \to0$ means that the corresponding limits hold both for $m^2\to0$ and $\tau\to0$. In this case, in the expansion \eqref{tildeEexp} the regular part dominates (and diverges as $\tau\to0$) in the relevant parameter region $\Re(\alpha-\mu)>0$, and the singular part dominates (and diverges as $m^2\to0$) in the irrelevant region $\Re(\alpha-\mu)<0$. Finally, to calculate the limit $\tau\to0$ for $\bbB_\alpha$ we must use the second representation in \eqref{CalEmu2}, i.e. the series $\calH_2^{\overline{\text{reg}}}(z)$ with respect to the variable $z = \sigma/2\tau^{1/\nu} \to\infty$, which leads to the following expression:
\begin{equation} \label{Lim21Eq}
\bbB_\alpha\!\Big[ \frac{e^{-\tau F^\nu}}{F^\mu} \Big| \sigma\Big]\xrightarrow[\tau\to0]{} \bbB_\alpha\!\big[ F^{-\mu} \big| \sigma\big].
\end{equation}

On the contrary, for $\nu<0$ the situation is exactly the opposite. Namely, using the expansions \eqref{CalEmu2prime} and \eqref{tildeEexpPrime} leads to the following three limits:
\begin{equation} \begin{aligned}
&\bbB_\alpha^\text{reg}\!\Big[ \frac{e^{-\tau F^\nu}}{F^\mu} \Big| \sigma \Big] \xrightarrow[\sigma, \tau \to0]{} \bbL_\alpha\!\Big[ \frac{e^{-\tau F^\nu}}{F^\mu} \Big], \\
&\bbB_\alpha^\text{UV}\!\Big[ \frac{e^{-\tau F^\nu}}{F^\mu} \Big| \sigma \Big] \xrightarrow[\sigma, \tau \to0]{} \bbB_\alpha\!\big[ F^{-\mu} \big| \sigma \big], \\
&\bbM_\alpha\!\Big[ \frac{e^{-\tau F^\nu}}{F^\mu} \Big| m^2\Big]\xrightarrow[\tau\to0]{} \bbM_\alpha\!\big[ F^{-\mu} \big| m^2\big].
\end{aligned} \end{equation}
The regular part in the basic kernel dominates (and diverges at $\tau\to0$) in the irrelevant region $\Re(\alpha-\mu)<0$, and the UV singular part dominates (and diverges at $\sigma\to0$) in the relevant region $\Re(\alpha-\mu)>0$.

\paragraph{The complete massive kernel.}

Let us make two quick remarks. First, the multiple MB integrals we are considering are functions of $N$ dimensionless combinations of $(N+1)$ dimensional parameters of the problem. Clearly, we can form these combinations in an infinite number of different ways, each of which will correspond to a different representation of the same kernel through different multiple MB integrals. However, we know that all these representations are equivalent, since they are related to each other in a simple way by the transformation formulas for the integrals \eqref{Transf1}-\eqref{Transf2}. Therefore, we only need to fix any of these representations (preferably the one that proves most convenient for some reason) and henceforth consider only it.

Second, in \cite{BKW25a}, we used tildes and subscripts to distinguish between different multiple MB integrals. This notation was not tied to any invariants of the functions under consideration and was therefore purely arbitrary and far from ideal. Although we use here without proof the integral representations obtained in \cite{BKW25a} and retain the general notation used there, it should be noted that the notations used here for specific functions do not correspond to those used there. The interested reader will easily restore the correspondence between them, using the transformations \eqref{Transf1}-\eqref{Transf2} if necessary.

For the complete kernel of interest to us at the moment, we have the following integral representation (see \cite{BKW25a}):
\begin{align}
&\bbW_\alpha\!\Big[ \frac{e^{-\tau F^\nu}}{F^\mu} \Big| \sigma, m^2\Big] = \frac{\tau^\frac{\mu-\alpha}{\nu}}{(4\pi)^{d/2}}\; H_1(\bm{z}), \label{tildeE1} \\
&\bm{z} = \left(\frac{\sigma}{2\tau^{1/\nu}}, m^2 \tau^{1/\nu}\right), \\
&h_1(\bm{s}) = \Gamma(s_1) \Gamma(s_2) \frac{\Gamma\left(\frac{s_2-s_1+\alpha-\mu}{\nu}\right)}{\nu\Gamma(s_2-s_1+\alpha)}. \label{tildeE3}
\end{align}

\begin{figure}
     \centering
     \begin{subfigure}[b]{0.34\textwidth}
         \centering
         \includegraphics[width=\textwidth]{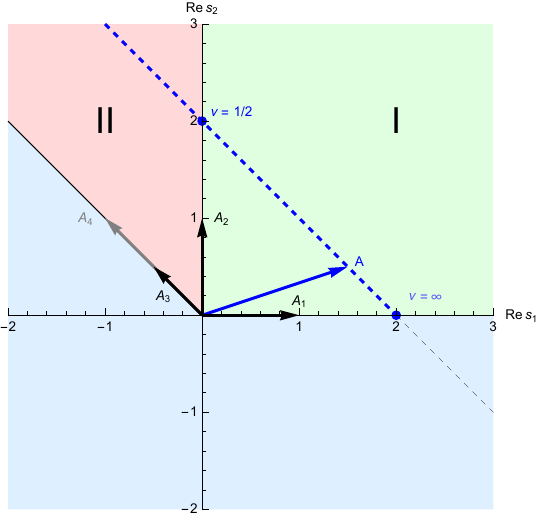}
         \caption{\footnotesize{The case $\nu>0$}}
         \label{Fig1a}
     \end{subfigure}
     \hfill
     \begin{subfigure}[b]{0.34\textwidth}
         \centering
         \includegraphics[width=\textwidth]{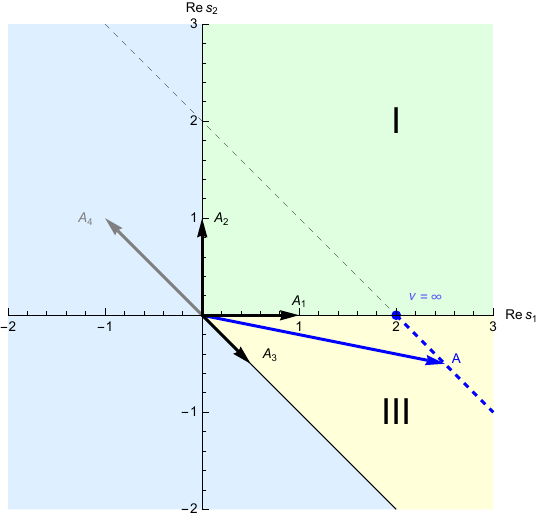}
         \caption{\footnotesize{The case $\nu<0$}}
         \label{Fig1b}
     \end{subfigure}
    \caption{\footnotesize{Vectors $\bm{A}_i$ and the corresponding elementary sectors for the function \eqref{tildeE3}. When the parameter $\nu$ changes, the vector $\bm{A}$ \eqref{bmAvector1} sweeps out a blue part of dashed line.}}
        \label{Fig1}
\end{figure}

In the notation of \eqref{GammaProduct}, the function $H_1(\bm{z})$ has the following parameters
\begin{equation} \label{Amatrix1} \begin{aligned}
\bm{\alpha} &= (1, 1, 1, -1), \\
\bm{a} &= (0, 0, \tfrac{\mu-\alpha}{\nu}, -\alpha), \\
A &= \begin{pmatrix}
1 & 0 & -1/\nu & -1 \\
0 & 1 & 1/\nu & 1
\end{pmatrix}.
\end{aligned} \end{equation}
The corresponding vectors $\bm{A}_i$ (i.e., columns of matrix $A$ \eqref{Amatrix1}) for the two cases $\nu>0$ and $\nu<0$ are schematically shown in Fig.~\ref{Fig1}. The resulting vector $\bm{A}$ \eqref{Param1}, which controls the convergence properties of the series, takes the form
\begin{equation} \label{bmAvector1}
\bm{A} = \binom{2-1/\nu}{1/\nu}.
\end{equation}
When changing the parameter $\nu$, it sweeps out the dashed line in Fig.~\ref{Fig1}. Note that it never becomes equal to zero, i.e. we are always dealing with the ``unbalanced'' case.

From here on, by $\{ij\}$ we mean the cluster generated by the vectors $\bm{A}_i$ and $\bm{A}_j$, and the corresponding cluster series will be denoted by $\calH_{ij}$. So three positive vectors $\bm{A}_1$, $\bm{A}_2$, and $\bm{A}_3$ determine three clusters: $\{12\}$, $\{13\}$, and $\{23\}$. These clusters will generate three families of simple poles, $n_i\in\bbZ_{\ge0}$:
\begin{equation}\begin{aligned}
&\bm{s}_{12}(\bm{n}) = (-n_1, -n_2), \\
&\bm{s}_{13}(\bm{n}) = (-n_1, \mu -\alpha - n_1 - \nu n_2), \\
&\bm{s}_{23}(\bm{n}) = (\alpha-\mu+\nu n_1 - n_2, -n_2).
\end{aligned}\end{equation}
Taking the appropriate residues, we obtain the following expressions for each of the series:
\begin{equation} \label{Hexpansion} \begin{aligned}
&\calH_{12}^\text{reg} = \sum\limits_{n_i = 0}^\infty\! \tfrac{\Gamma\left(\frac{\alpha-\mu+n_1-n_2}{\nu}\right)}{\nu\Gamma(\alpha+n_1-n_2)} \tfrac{(-z_1)^{n_1}}{n_1!} \tfrac{(-z_2)^{n_2}}{n_2!}, \\
&\calH_{13} = z_2^{\alpha-\mu} \!\sum\limits_{n_i = 0}^\infty\! \tfrac{\Gamma(\mu-\alpha-n_1-\nu n_2)}{\Gamma(\mu - \nu n_2)} \tfrac{(-z_1z_2)^{n_1}}{n_1!} \tfrac{(-z_2^\nu)^{n_2}}{n_2!}, \\
&\calH_{23} = -z_1^{\mu-\alpha} \!\sum\limits_{n_i = 0}^\infty\! \tfrac{\Gamma(\alpha-\mu+\nu n_1-n_2)}{\Gamma(\mu - \nu n_1)} \tfrac{(-z_1^{-\nu})^{n_1}}{n_1!} \tfrac{(-z_1z_2)^{n_2}}{n_2!}.
\end{aligned} \end{equation}

Note that, in the non-resonant case of arbitrary $\alpha$ and $\mu=0$, $\nu = 2,3,4,\ldots$, cluster series $\calH_{13}$ and $\calH_{23}$ vanish identically. This is because in the expression \eqref{tildeE3} all poles of the function $\Gamma\big((s_2-s_1+\alpha-\mu)/\nu\big)$ in the numerator are actually canceled by the poles of the function $\Gamma(s_2-s_1+\alpha)$ in the denominator, and thus the vector $\bm{A}_3$ does not generate any singularities at all. Then the function $H_1(\bm{z})$ is reduced to a single cluster series $\calH_{12}^\text{reg}$. This phenomenon is completely analogous to the case of exponential asymptotics of $\calE_{\nu,\alpha}(z)$ for positive integers $\nu$, which we considered in detail in \cite{Wach2}.

In the two cases we are considering, $\nu>0$ and $\nu<0$, the cluster series \eqref{Hexpansion} will be the same, but the arrangement of the elementary sectors will be different. Let us consider each of these cases in turn.

\paragraph{The cases $\nu>0$ and $\nu<0$.}

When $\nu>0$, we have two elementary sectors, shown in Fig.~\ref{Fig1a}: green sector I, lying between vectors $\bm{A}_1$ and $\bm{A}_2$, $\varphi\in(0,\pi/2)$, and pink sector II, lying between vectors $\bm{A}_2$ and $\bm{A}_3$, $\varphi\in(\pi/2, 3\pi/4)$. They correspond to two cluster representations:
\begin{equation} \label{ClusterH}
H_1(\bm{z}) = \calH_{12}^\text{reg} + \calH_{13}^\text{IR} = \calH_{23}^{\overline{\text{reg}}} + \calH_{13}^\text{IR}.
\end{equation}
The cluster series $\calH_{13}^\text{IR}$ is everywhere convergent for any $\nu>0$. For $\nu>1/2$, the vector $\bm{A}$ \eqref{bmAvector1} lies in sector I, therefore the series $\calH_{12}^\text{reg}$ also converges everywhere, and the series $\calH_{23}^{\overline{\text{reg}}}$ is asymptotic. For $0<\nu<1/2$, the vector $\bm{A}$ \eqref{bmAvector1} lies in sector II, so the situation is the opposite: the series $\calH_{23}^{\overline{\text{reg}}}$ converges everywhere, while $\calH_{12}^\text{reg}$ is asymptotic.

In the expansion \eqref{ClusterH}, we interpret the cluster series $\calH_{12}^\text{reg}$ and $\calH_{23}^{\overline{\text{reg}}}$ as two different representations for the regular part of the complete massive kernel \eqref{tildeE1}, and the series $\calH_{13}^\text{IR}$ as the corresponding IR singular contribution. This interpretation is confirmed by a direct verification of our fundamental expansions \eqref{FundExpansion2}-\eqref{FundExpansion3} using the previously obtained cluster representations \eqref{CalEmu2} and \eqref{tildeEexp}. Accordingly, the limits \eqref{FundLimits} automatically hold.

When $\nu<0$, the vector $\bm{A}_3$ changes its direction and we obtain new elementary sectors, shown in Fig.~\ref{Fig1b}: green sector I, lying between the vectors $\bm{A}_1$ and $\bm{A}_2$, $\varphi\in(0,\pi/2)$, and yellow sector III, lying between the vectors $\bm{A}_1$ and $\bm{A}_3$, $\varphi\in(-\pi/4, 0)$. They correspond to two cluster representations:
\begin{equation} \label{ClusterHprime}
H_1(\bm{z}) = \calH_{12}^\text{reg} + \calH_{23}^\text{UV} = \calH_{13}^{\overline{\text{reg}}} + \calH_{23}^\text{UV}.
\end{equation}
The vector $\bm{A}$ \eqref{bmAvector1} lies in sector III for any value of $\nu<0$. Therefore, the cluster series $\calH_{13}^{\overline{\text{reg}}}$ and $\calH_{23}^\text{UV}$ are everywhere convergent, and the series $\calH_{12}^\text{reg}$ is asymptotic.

Accordingly, the interpretation of our cluster series will also be different: the series $\calH_{13}^{\overline{\text{reg}}}$ now defines a different representation of the regular part, and the series $\calH_{23}^\text{UV}$ corresponds to the UV singular contribution. This interpretation is again confirmed by a direct test of the fundamental expansions \eqref{FundExpansion1} and \eqref{FundExpansion3}, this time using different cluster representations \eqref{CalEmu2prime} and \eqref{tildeEexpPrime}.

\paragraph{For the limit $\tau\to0$,} we have
\begin{equation} \label{AppEq6}
z_1 \to\infty, \quad z_2 \to0,  \quad z = z_1z_2 = \frac{\sigma m^2}{2} = \mathrm{const}.
\end{equation}
Accordingly, to calculate it, we need to use the second representation in \eqref{ClusterH} or in \eqref{ClusterHprime}. These representations coincide in terms of the cluster series they contain, and differ only in their interpretation: the series $\calH_{23}$ is interpreted as regular for $\nu>0$ and as UV singular for $\nu<0$, while the series $\calH_{13}$ is interpreted as IR singular for $\nu>0$ and as regular for $\nu<0$. But one way or another, regardless of the sign of $\nu$, we obtain the same limit:
\begin{equation}
\bbW_\alpha\!\Big[ \frac{e^{-\tau F^\nu}}{F^\mu} \Big| \sigma, m^2\Big] \xrightarrow[\tau\to0]{} \bbW_\alpha\big[F^{-\mu} \big| \sigma, m^2 \big],
\end{equation}
Moreover, it is easy to verify that, regardless of the interpretation of the corresponding cluster series, the UV singular part of $\bbW_\alpha\big[F^{-\mu} \big| \sigma, m^2 \big]$ arises from the series $\calH_{23}$, and the IR singular part arises from the series $\calH_{13}$.

To summarize, the obtained results lead us to the following natural conjecture: singular contributions are determined by the power-law asymptotic behavior of the operator function $f(\hat F)$ in the UV and IR limits. If the operator function $f(\hat F)$ decreases exponentially in this limit, i.e., faster than any power, the corresponding singular contribution is absent.

\subsection{The resolvent $(\hat F^\mu + \lambda)^{-1}$} \label{ResolventSubsec}

\paragraph{Basis and complimentary kernels.}

We will use the following integral representation (see \cite{BKW25a}, the parameter $\mu\ne0$ is no longer complex, but real) for the basic kernel and complementary kernels :
\begin{align}
&\bbB_\alpha\!\Big[\frac{1}{F^\mu + \lambda} \Big| \sigma\Big] = \frac{\lambda^{\frac{\alpha}{\mu}-1}}{(4\pi)^{d/2}}\; \calG_{\mu,\alpha}(z), \quad z = \frac{\sigma\lambda^{1/\mu}}{2}, \label{calGEq1} \\
&g_{\mu,\alpha}(s) = \frac{\Gamma(s)}{\mu\Gamma(\alpha-s)}\,\Gamma\!\left(\frac{\alpha-s}{\mu}\right)\,\Gamma\!\left(1-\frac{\alpha-s}{\mu}\right), \label{calGEq3}
\end{align}
and
\begin{align}
&\bbM_\alpha\!\Big[\frac{1}{F^\mu + \lambda} \Big| m^2\Big] = \frac{\lambda^{\frac{\alpha}{\mu}-1}}{(4\pi)^{d/2}}\; \tilde\calG_{\mu,\alpha}(z), \quad z = \frac{m^2}{\lambda^{1/\mu}}, \label{calGEq2} \\
&\tilde g_{\mu,\alpha}(s) = \frac{\Gamma(s)}{\mu\Gamma(\alpha+s)}\,\Gamma\!\left(\frac{\alpha+s}{\mu}\right)\,\Gamma\!\left(1-\frac{\alpha+s}{\mu}\right). \label{calGEq4}
\end{align}

It is easy to see that the difference between these two functions is only in the sign of the argument of the gamma factor in the denominator of $g_{\mu,\alpha}(s)$ and $\tilde g_{\mu,\alpha}(s)$, while the pole structure is similar: in both cases, there is a left-running row of poles $s_n = -n$ and a bidirectional row (we consider the non-resonant case, i.e., we assume throughout that the parameters $\mu$ and $\alpha$ are such that these poles do not coalesce). Since this structure obviously does not change when $\mu$ changes sign, for definiteness we will henceforth assume that $\mu>0$.

The bidirectional row must be split into a leftward-running row and a rightward-running row, and the exact location of this splitting plays an important role, since it determines the integration contour and the resulting function. This is precisely why, in formulas \eqref{calGEq3}-\eqref{calGEq4}, we write out all the gamma factors explicitly, rather than hiding two of them into an inverse sine using the reflection formula \eqref{ReflectionFormula}---this encodes the correct splitting. Further, by closing the integration loop to the left, in both cases we obtain an expansion of the functions near $z=0$ as the sum of two series, and by closing the loop to the right, we obtain an expansion near $z=\infty$ as a single series.

Thus, for the function $g_{\mu,\alpha}(s)$ \eqref{calGEq3} we have three clusters generating three rows of poles: $s_n = -n$ and $s_k = \alpha \pm \mu k$, where $n\in\bbZ_{\ge0}$ and $k\in\bbZ_{\ge1}$. The sums of residues at the poles from these three rows gives three cluster series:
\begin{equation} \label{Gexpansion} \begin{aligned}
&\calH_1^\text{reg}(z) = \sum\limits_{n=0}^\infty \frac{\pi/\mu}{\sin\tfrac{\pi}{\mu}(\alpha+n)} \frac{(-z)^n}{\Gamma(\alpha+n) n!},  \\
&\calH_2^\text{UV}(z) = - z^{-\alpha} \sum\limits_{k=1}^\infty \frac{\Gamma(\alpha - \mu k)}{\Gamma(\mu k)} (-z^\mu)^k, \\
&\calH_3^\text{reg+UV}(z) = -z^{-\alpha} \sum\limits_{k=1}^\infty \frac{\Gamma(\alpha + \mu k)}{\Gamma(-\mu k)} (-z^{-\mu})^k.
\end{aligned} \end{equation}
Similarly, for the function $\tilde g_{\mu,\alpha}(s)$ \eqref{calGEq4} we also have three clusters, generating three rows of poles: $s_n = -n$, $n\in\bbZ_{\geq0}$ and $s_k = -\alpha \pm \mu k$, $k\in\bbZ_{\geq1}$. The sums of residues at these poles yields three other cluster series:
\begin{equation} \label{tildeGexpansion} \begin{aligned}
&\tilde\calH_1^\text{reg}(z) = \sum\limits_{n=0}^\infty \frac{\pi/\mu}{\sin\tfrac{\pi}{\mu}(\alpha-n)} \frac{(-z)^n}{\Gamma(\alpha-n) n!},  \\
&\tilde\calH_2^\text{IR}(z) = z^\alpha \sum\limits_{k=1}^\infty \frac{\Gamma(-\alpha - \mu k)}{\Gamma(-\mu k)} (-z^\mu)^k, \\
&\tilde\calH_3^\text{reg+IR}(z) = z^\alpha \sum\limits_{k=1}^\infty \frac{\Gamma(-\alpha + \mu k)}{\Gamma(\mu k)} (-z^{-\mu})^k.
\end{aligned} \end{equation}

Let us write out the corresponding representations as sums over clusters:
\begin{equation} \label{Gexpansions} \begin{aligned} 
\calG_{\mu,\alpha}(z) &= \calH_1^\text{reg}(z) + \calH_2^\text{UV}(z) = \calH_3^\text{reg+UV}(z), \\
\tilde\calG_{\mu,\alpha}(z) &= \tilde\calH_1^\text{reg}(z) + \tilde\calH_2^\text{IR}(z) = \tilde\calH_3^\text{reg+IR}(z).
\end{aligned} \end{equation}

The cluster series \eqref{Gexpansion}-\eqref{tildeGexpansion} participating in these expansions have the following convergence properties. For the function $\calG_{\mu,\alpha}(z)$ we have an unbalanced case: the series $\calH_1^\text{reg}(z)$ and $\calH_2^\text{UV}(z)$ near $z=0$ are convergent everywhere for all parameter values, while the series $\calH_3^\text{reg+UV}(z)$ near $z=\infty$ is asymptotic. Conversely, for the function $\tilde\calG_{\mu,\alpha}(z)$, we always have a balanced case: the series $\tilde\calH_1^\text{reg}(z)$ and $\tilde\calH_2^\text{IR}(z)$ near the point $z=0$ will converge inside the disk $|z|<1$ and diverge outside it, while the series $\tilde\calH_3^\text{reg+IR}(z)$ near the point $z=\infty$ will do the opposite way.

Above, we denoted the interpretations of the corresponding cluster series by superscripts. The series $\calH_1^\text{reg}(z)$ and $\tilde\calH_1^\text{reg}(z)$ represent the regular contributions, and the series $\calH_2^\text{UV}(z)$ and $\tilde\calH_2^\text{IR}(z)$ are the corresponding singular contributions. A new phenomenon here is the appearance of the series $\calH_3^\text{reg+UV}(z)$ and $\tilde\calH_3^\text{reg+IR}(z)$, which represent the sum of the regular and singular contributions in the opposite region of the arguments.

If we compare the series \eqref{Gexpansion} and \eqref{tildeGexpansion} for the regular contributions, we easily obtain the following expression for the regular kernel
\begin{equation} \label{ResolventRegularKernel}
\bbL_\alpha\!\big[ (F^\mu + \lambda)^{-1} \big] = \frac{\pi/\mu}{\sin\tfrac{\pi\alpha}{\mu}} \frac{\lambda^{\frac{\alpha}{\mu}-1}}{(4\pi)^{d/2}\Gamma(\alpha)},
\end{equation}
such that two limits in \eqref{FundLimits} hold.

Note again that when we calculate the coincidence limit $\sigma\to0$ for the regular part of the basis kernel, we use the series $\calH_1^\text{reg}$ in the variable $z = \sigma\lambda^{1/\mu}/2$. Therefore, for $\lambda\to0$ we obtain the same limit. But the same will be true for the limits of the UV singular part, in this case we should only use the series $\calH_2^\text{UV}$. As a result, we have:
\begin{equation} \begin{aligned}
&\bbB_\alpha^\text{reg}\!\big[ (F^\mu + \lambda)^{-1} \big| \sigma\big] \xrightarrow[\sigma, \lambda\to0]{} \bbL_\alpha\!\big[ (F^\mu + \lambda)^{-1} \big], \\
&\bbB_\alpha^\text{UV}\!\big[ (F^\mu + \lambda)^{-1} \big| \sigma\big] \xrightarrow[\sigma, \lambda\to0]{} \bbB_\alpha\!\big[ F^{-\mu} \big| \sigma\big].
\end{aligned} \end{equation}
As usual, the regular contribution will dominate in the irrelevant region $\Re(\alpha-\mu)<0$, and the UV singular contribution will dominate in the relevant region $\Re(\alpha-\mu)>0$.

To calculate the limit $\lambda\to0$, we must use the series $\tilde\calH_3^\text{reg+IR}(z)$, which immediately gives us
\begin{equation}
\bbM_\alpha\!\big[ (F^\mu + \lambda)^{-1} \big| m^2 \big] \xrightarrow[\lambda\to0]{} \bbM_\alpha\!\big[ F^{-\mu} \big| m^2 \big].
\end{equation}

\paragraph{The complete massive kernel} can be written as the following integral representation (see \cite{BKW25a}):
\begin{align}
&\bbW_\alpha\!\Big[ \frac{1}{F^\mu + \lambda} \Big| \sigma, m^2\Big] = \frac{\lambda^{\frac{\alpha}{\mu}-1}}{(4\pi)^{d/2}}\; H_2(\bm{z}), \label{AppEq12} \\
&\bm{z} = \left(\frac{\sigma\lambda^{1/\mu}}{2}, \frac{m^2}{\lambda^{1/\mu}}\right), \\
&h_2(\bm{s}) = -\Gamma(s_1)\,\Gamma(s_2)\,\frac{\Gamma\!\left(\tfrac{\alpha+s_2-s_1}{\mu}\right)\, \Gamma\!\left(1 - \tfrac{\alpha+s_2-s_1}{\mu}\right)}{\mu \Gamma(\alpha+s_2-s_1)}. \label{FunctionH2}
\end{align}

\begin{figure}
     \centering
     \begin{subfigure}[b]{0.34\textwidth}
         \centering
         \includegraphics[width=\textwidth]{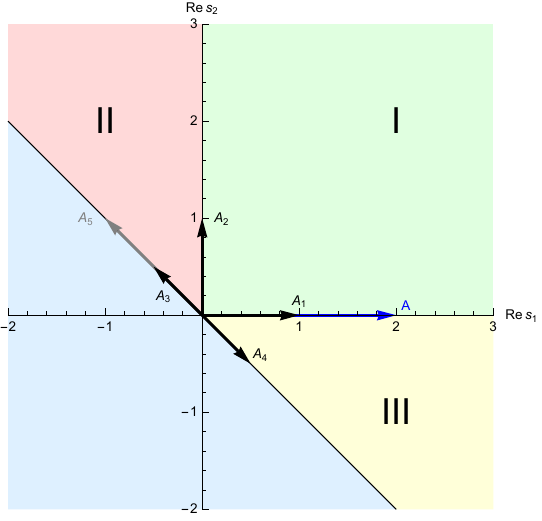}
         \caption{\footnotesize{For the function $h_2(\bm{s})$ \eqref{FunctionH2}.}}
         \label{Fig2a}
     \end{subfigure}
     \hfill
     \begin{subfigure}[b]{0.34\textwidth}
         \centering
         \includegraphics[width=\textwidth]{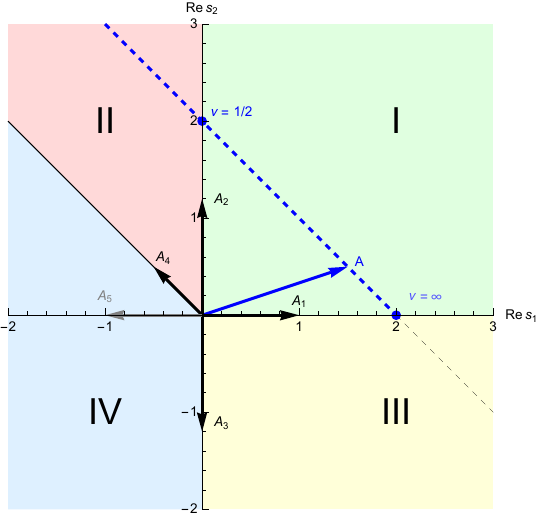}
         \caption{\footnotesize{For the function $h_3(\bm{s})$ \eqref{HibridBasis3}.}}
         \label{Fig2b}
     \end{subfigure}
\hfill
     \begin{subfigure}[b]{0.34\textwidth}
         \centering
         \includegraphics[width=\textwidth]{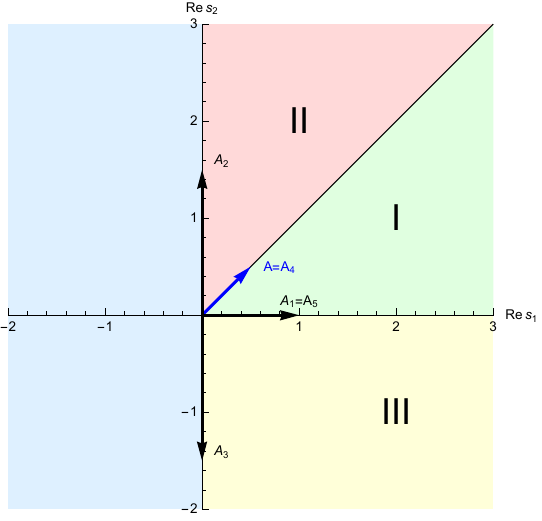}
         \caption{\footnotesize{For the function $\tilde h_3(\bm{s})$ \eqref{tildeHibridBasis3}.}}
         \label{Fig2c}
     \end{subfigure}
    \caption{\footnotesize{Vectors $\bm{A}_i$ and the corresponding elementary sectors for different 2-fold MB integrals.}}
        \label{Fig2}
\end{figure}

Again following the notation \eqref{GammaProduct}, we write out the parameters for the function $H_2(\bm{z})$:
\begin{equation} \label{Amatrix2} \begin{aligned}
\bm{\alpha} &= (1, 1, 1, 1, -1), \\
\bm{a} &= \left(0, 0, -\tfrac{\alpha}{\mu}, \tfrac{\alpha}{\mu}-1, -\alpha\right), \\
A &= \begin{pmatrix}
1 & 0 & -1/\mu & 1/\mu & -1 \\
0 & 1 & 1/\mu & -1/\mu & 1
\end{pmatrix}.
\end{aligned}\end{equation}
The corresponding vectors $\bm{A}_i$ and the elementary sectors they define are schematically depicted in Fig.~\ref{Fig2a}. As above, for definiteness, we assume throughout that $\mu>0$.

There are five clusters $\{12\}, \{13\}, \{14\}, \{23\}, \{24\}$, which generate the following families of simple poles, $n\in\bbZ_{\geq0}$, $n_i\in\bbZ_{\ge0}$, and $k\in\bbZ_{\ge1}$:
\begin{equation}\begin{aligned}
&\bm{s}_{12}(n_1, n_2) = (-n_1, -n_2), \\
&\bm{s}_{13}(n, k) = (-n, -\alpha-\mu k-n), \\
&\bm{s}_{14}(n, k) = (-n, -\alpha+\mu k-n), \\
&\bm{s}_{23}(n, k) = (\alpha+\mu k-n, -n), \\
&\bm{s}_{24}(n, k) = (\alpha-\mu k-n, -n).
\end{aligned} \end{equation}
The sums of the residues at these poles generate the following cluster series, $n\in\bbZ_{\ge0}$ and $k\in\bbZ_{\ge1}$:
\begin{equation} \label{ResolvSeries} \begin{aligned}
&\calH_{12}^\text{reg} = -\sum\limits_{n_1, n_2} \tfrac{\pi \Gamma^{-1}(\alpha+n_1-n_2)}{\mu\sin\tfrac{\pi}{\mu}(\alpha+n_1-n_2)} \tfrac{(-z_1)^{n_1}}{n_1!} \tfrac{(-z_2)^{n_2}}{n_2!}, \\
&\calH_{13}^\text{IR} = -z_2^\alpha \sum\limits_{n, k} \tfrac{\Gamma(-\alpha-\mu k-n)}{\Gamma(-\mu k)} \tfrac{(-z_1z_2)^n}{n!} (-z_2^\mu)^k, \\
&\calH_{14}^\text{reg+IR} = -z_2^\alpha \sum\limits_{n, k} \tfrac{\Gamma(-\alpha+\mu k-n)}{\Gamma(\mu k)} \tfrac{(-z_1z_2)^n}{n!} (-z_2^{-\mu})^k, \\
&\calH_{23}^\text{reg+UV} = z_1^{-\alpha} \sum\limits_{n, k} \tfrac{\Gamma(\alpha+\mu k-n)}{\Gamma(-\mu k)} \tfrac{(-z_1z_2)^n}{n!} (-z_1^{-\mu})^k, \\
&\calH_{24}^\text{UV} = z_1^{-\alpha} \sum\limits_{n, k} \tfrac{\Gamma(\alpha-\mu k-n)}{\Gamma(\mu k)} \tfrac{(-z_1z_2)^n}{n!} (-z_1^\mu)^k.
\end{aligned}\end{equation}

The interpretation of these series, which we indicate with the help of superscripts, can be confirmed by a direct check of our fundamental expansions \eqref{FundExpansion1}-\eqref{FundExpansion3} and the limits \eqref{FundLimits} that follow from them, using the previously obtained expansions \eqref{Gexpansion}-\eqref{tildeGexpansion}. But it can also be seen more simply from the way our variables $z_1 = \sigma\lambda^{1/\mu}/2$ and $z_2 = m^2/\lambda^{1/\mu}$ enter into the series under consideration. For example, the IR singular terms should be well defined in the coincidence limit $z_1\to0$, $z_2 = \mathrm{const}$ and poorly defined in the massless limit $z_2\to0$, $z_1 = \mathrm{const}$, and the series $\calH_{13}^\text{IR}$ and $\calH_{14}^\text{reg+IR}$ have such properties.

Further, we have three elementary sectors, shown in Fig.~\ref{Fig2a}: the green sector I, lying between the vectors $\bm{A}_1$ and $\bm{A}_2$, $\varphi\in(0, \pi/2)$; the pink sector II, lying between the vectors $\bm{A}_2$ and $\bm{A}_3$, $\varphi\in(\pi/2, 3\pi/4)$; yellow sector III, lying between the vectors $\bm{A}_1$ and $\bm{A}_4$, $\varphi\in(-\pi/4, 0)$. According to the formula \eqref{SumOverClusters}, these three sectors correspond to three different representations of the function as an expansion in cluster series:
\begin{align}
H_2(\bm{z}) &= \calH_{12}^\text{reg} + \calH_{24}^\text{UV} + \calH_{13}^\text{IR} \nonumber \\
&= \calH_{24}^\text{UV} + \calH_{14}^\text{reg+IR} = \calH_{13}^\text{IR} + \calH_{23}^\text{reg+UV}. \label{GclusterSums}
\end{align}

The resulting vector \eqref{Param1}, which determines the convergence properties of the cluster series, has the form
\begin{equation} \label{bmAvector2}
\bm{A} = \binom{2}{0}.
\end{equation}
It is independent of the parameter $\mu$ and lies exactly on the boundary between elementary sectors I and III. It follows that the series $\calH_{24}^\text{UV}$ always converges; the series $\calH_{23}^\text{reg+UV}$ is always asymptotic; finally, the series $\calH_{12}^\text{reg}$ and $\calH_{13}^\text{IR}$ converge in a certain region around the plane $z_2=0$ and diverge outside it, while the series $\calH_{14}^\text{reg+IR}$ behaves in the opposite way.

Finally, we calculate the limit of $\lambda\to0$, i.e. $z_1\to0$, $z_2\to\infty$, and $z = z_1z_2 = \sigma m^2/2 = \mathrm{const}$. As can be seen from the series \eqref{ResolvSeries}, to calculate it we need to use the second representation in \eqref{GclusterSums}, associated with the elementary sector III. Substituting the corresponding expansions into \eqref{AppEq12} and comparing the answer with \eqref{GreenIRregApp}, we find
\begin{equation}
\bbW_\alpha\big[(F^\mu + \lambda)^{-1} \big| \sigma, m^2 \big] \xrightarrow[\lambda\to0]{} \bbW_\alpha\big[F^{-\mu} \big| \sigma, m^2 \big].
\end{equation}
In this case, the UV singular part of $\bbW_\alpha\big[F^{-\mu} \big| \sigma, m^2 \big]$ arises from the series $\calH_{24}^\text{UV}$, and the IR singular part comes from the series $\calH_{14}^\text{reg+IR}$.

\subsection{The function $e^{-\tau\hat F^\nu}/(\hat F^\mu + \lambda)$} \label{TheMostDifficultSubsec}

Now let us consider the most complex of our examples. To avoid overloading the reader, we will not dwell on previously encountered details, but will present only the main results, emphasizing new phenomena. For brevity and simplicity, we will assume throughout that $\nu>0$ and $\mu>0$.

The interpretation we introduced in Sec.~\ref{StructureSec}, in which we distinguish between the regular, UV, and IR contributions, is very convenient in that we can unambiguously identify and compare different cluster series. However, in current more complex case, more cluster series and representations constructed from them arise. Therefore, our initial interpretation is no longer sufficient for these purposes and requires some extension. Specifically, among the regular terms, we distinguish contributions corresponding to each of our dimensional parameters $\tau$ and $\lambda$, as giving a nontrivial contribution in the limit when we let the other parameter tend to zero. We will continue to denote these $\tau$- and $\lambda$-contributions by superscripts.

\paragraph{The basis kernel} can be written as the following integral representation:
\begin{align}
&\bbB_\alpha\!\Big[ \frac{e^{-\tau F^\nu}}{F^\mu + \lambda} \Big| \sigma \Big] = \frac{\tau^\frac{\mu-\alpha}{\nu}}{(4\pi)^{d/2}}\; H_3(\bm{z}), \label{HibridBasis1} \\
&\bm{z} = \left(\frac{\sigma}{2\tau^{1/\nu}}, \lambda^{1/\mu} \tau^{1/\nu}\right), \\
&h_3(\bm{s}) = \frac{\Gamma(s_1)\, \Gamma(\tfrac{s_2}{\mu})\, \Gamma(1-\tfrac{s_2}{\mu})}{\mu\nu\Gamma(\alpha - s_1)}\, \Gamma\!\left(\tfrac{s_2 - s_1 + \alpha - \mu}{\nu}\right). \label{HibridBasis3}
\end{align}

The parameters of the function $H_3(\bm{z})$ have the form
\begin{equation} \label{Amatrix3} \begin{aligned}
\bm{\alpha} &= (1, 1, 1, 1, -1), \\
\bm{a} &= (0, 0, -1, \tfrac{\mu-\alpha}{\nu}, -\alpha), \\
A &= \begin{pmatrix}
1 & 0 & 0 & -1/\nu & -1 \\
0 & 1/\mu & -1/\mu & 1/\nu & 0
\end{pmatrix}.
\end{aligned} \end{equation}
The corresponding vectors $\bm{A}_i$ and the elementary sectors generated by them are schematically depicted in Fig.~\ref{Fig2b}.

We have five clusters: $\{12\}, \{13\}, \{14\}, \{24\}, \{34\}$. They generate the following families of simple poles, $n_i\in\bbZ_{\ge0}$:
\begin{equation} \label{BasisPolesFamilies} \begin{aligned}
&\bm{s}_{12}(\bm{n}) = \big(-n_1, -\mu n_2\big), \\
&\bm{s}_{13}(\bm{n}) = \big(-n_1, \mu(n_2+1)\big), \\
&\bm{s}_{14}(\bm{n}) = \big(-n_1, \mu-\alpha-n_1-\nu n_2\big), \\
&\bm{s}_{24}(\bm{n}) = \big(\alpha + \nu n_1 - \mu (n_2+1), -\mu n_2\big), \\
&\bm{s}_{34}(\bm{n}) = \big(\alpha + \nu n_1 + \mu n_2, \mu(n_2+1)\big).
\end{aligned}\end{equation}

The sums of residues at these poles give the following cluster series:
\begin{equation} \label{ClSeries1} \begin{aligned}
&\calH_{12}^\tau = \sum\limits_{n_i = 0}^\infty \frac{\Gamma\!\left(\tfrac{\alpha + n_1 - \mu(n_2+1)}{\nu}\right)}{\nu\Gamma(\alpha + n_1)} \frac{(-z_1)^{n_1}}{n_1!} (-z_2^\mu)^{n_2}, \\
&\calH_{13}^{\tau + \lambda} = -z_2^{-\mu} \sum\limits_{n_i = 0}^\infty \frac{\Gamma\!\left(\tfrac{\alpha + n_1 + \mu n_2}{\nu}\right)}{\nu\Gamma(\alpha + n_1)} \frac{(-z_1)^{n_1}}{n_1!} (-z_2^{-\mu})^{n_2}, \\
&\calH_{14}^\lambda = z_2^{\alpha-\mu} \sum\limits_{n_i = 0}^\infty \tfrac{\pi \Gamma^{-1}(\alpha + n_1) / \mu}{\sin\left(\tfrac{\pi}{\mu}(\alpha+n_1+\nu n_2)\right)} \tfrac{(-z_1z_2)^{n_1}}{n_1!} \tfrac{(-z_2^\nu)^{n_2}}{n_2!}, \\
&\calH_{24}^{\overline{\tau}} = -z_1^{\mu-\alpha} \sum\limits_{n_i = 0}^\infty \tfrac{\Gamma\big(\alpha + \nu n_1 - \mu(n_2+1)\big)}{\Gamma\big(\mu(n_2+1) - \nu n_1\big)} \tfrac{(-z_1^{-\nu})^{n_1}}{n_1!} (-z_1^\mu z_2^\mu)^{n_2}, \\
&\calH_{34}^{\overline{\tau + \lambda}} = z_1^{-\alpha}z_2^{-\mu} \sum\limits_{n_i = 0}^\infty \tfrac{\Gamma(\alpha + \nu n_1 + \mu n_2)}{\Gamma(-\mu n_2 - \nu n_1)} \tfrac{(-z_1^{-\nu})^{n_1}}{n_1!} (-z_1^{-\mu} z_2^{-\mu})^{n_2}.
\end{aligned}\end{equation}

Next, we have four elementary sectors, shown in different colors in Fig.~\ref{Fig2b}. They correspond to four representations in the form of sums over clusters:
\begin{align}
H_3(\bm{z}) &= \calH_{12}^\tau + \calH_{14}^\lambda = \calH_{13}^{\tau + \lambda} \nonumber \\
&= \calH_{24}^{\overline{\tau}} + \calH_{14}^{\lambda} = \calH_{34}^{\overline{\tau + \lambda}}. \label{BasisRepresentations}
\end{align}

The vector $\bm{A}$ \eqref{Param1} again takes the form \eqref{bmAvector1}. Therefore, the cluster series $\calH^\lambda_{14}$ is everywhere convergent, while the series $\calH^{\tau+\lambda}_{13}$ and $\calH^{\overline{\tau+\lambda}}_{34}$ are asymptotic. For $1/2<\nu<\infty$, when the vector $\bm{A}$ lies in sector I, the series $\calH^\tau_{12}$ converges everywhere, while the series $\calH^{\overline{\tau}}_{24}$ is asymptotic, and for $0<\nu<1/2$, when the vector $\bm{A}$ moves to sector II, the situation is reversed.

\paragraph{The complementary kernel} can be written as the following integral representation:
\begin{align}
&\bbM_\alpha\!\Big[ \frac{e^{-\tau F^\nu}}{F^\mu + \lambda} \Big| m^2 \Big] = \frac{\tau^\frac{\mu-\alpha}{\nu}}{(4\pi)^{d/2}}\; \tilde H_3(\bm{z}), \label{tildeHibridBasis1} \\
&\bm{z} = \left(m^2\tau^{1/\nu}, \lambda^{1/\mu} \tau^{1/\nu}\right), \\
&\tilde h_3(\bm{s}) = \frac{\Gamma(s_1)\, \Gamma(\tfrac{s_2}{\mu})\, \Gamma(1-\tfrac{s_2}{\mu})}{\mu\nu\Gamma(\alpha + s_1)}\, \Gamma\!\left(\tfrac{s_1 + s_2 + \alpha - \mu}{\nu}\right). \label{tildeHibridBasis3}
\end{align}

For the function $\tilde H_3(\bm{z})$, the parameters $\bm{\alpha}$ and $\bm{a}$ look the same as in \eqref{Amatrix3}, and the matrix $A$ has the form
\begin{equation} \label{tildeAmatrix3}
A = \begin{pmatrix}
1 & 0 & 0 & 1/\nu & 1 \\
0 & 1/\mu & -1/\mu & 1/\nu & 0
\end{pmatrix}. \end{equation}
The corresponding vectors $\bm{A}_i$ and the elementary sectors they generate are schematically shown in Fig.~\ref{Fig2c}.

As in the case of the basic kernel, we will have five clusters: $\{12\}, \{13\}, \{14\}, \{24\}, \{34\}$. The pole families generated by the clusters $\{12\}$ and $\{13\}$ will be the same as in \eqref{BasisPolesFamilies}: $\tilde{\bm s}_{12}(\bm{n}) = \bm{s}_{12}(\bm{n})$, $\tilde{\bm s}_{13}(\bm{n}) = \bm{s}_{13}(\bm{n})$. And the families of poles generated by the three remaining clusters will be slightly different:
\begin{equation}\begin{aligned}
&\tilde{\bm s}_{14}(\bm{n}) = \big(-n_1, \mu-\alpha+n_1-\nu n_2\big), \\
&\tilde{\bm s}_{24}(\bm{n}) = \big(-\alpha - \nu n_1 + \mu (n_2+1), -\mu n_2\big), \\
&\tilde{\bm s}_{34}(\bm{n}) = \big(-\alpha - \nu n_1 - \mu n_2, \mu(n_2+1)\big).
\end{aligned}\end{equation}

The sums of the residues at these poles yield the following cluster series:
\begin{equation} \label{ClSeries2} \begin{aligned}
&\tilde\calH_{12}^\tau = \sum\limits_{n_i = 0}^\infty \frac{\Gamma\!\left(\tfrac{\alpha - n_1 - \mu(n_2+1)}{\nu}\right)}{\nu\Gamma(\alpha - n_1)} \frac{(-z_1)^{n_1}}{n_1!} (-z_2^\mu)^{n_2}, \\
&\tilde\calH_{13}^{\tau + \lambda} = -z_2^{-\mu} \sum\limits_{n_i = 0}^\infty \frac{\Gamma\!\left(\tfrac{\alpha - n_1 + \mu n_2}{\nu}\right)}{\nu\Gamma(\alpha - n_1)} \frac{(-z_1)^{n_1}}{n_1!} (-z_2^{-\mu})^{n_2}, \\
&\tilde\calH_{14}^\lambda = z_2^{\alpha-\mu} \sum\limits_{n_i = 0}^\infty \tfrac{\pi \Gamma^{-1}(\alpha - n_1) / \mu}{\sin\left(\tfrac{\pi}{\mu}(\alpha-n_1+\nu n_2)\right)} \tfrac{(-z_1/z_2)^{n_1}}{n_1!} \tfrac{(-z_2^\nu)^{n_2}}{n_2!}, \\
&\tilde\calH_{24}^{\lambda+\text{IR}} = z_1^{\alpha-\mu} \sum\limits_{n_i = 0}^\infty \tfrac{\Gamma\big(\mu(n_2+1) - \nu n_1 - \alpha\big)}{\Gamma\big(\mu(n_2+1) - \nu n_1\big)} \tfrac{(-z_1^\nu)^{n_1}}{n_1!} (-z_2^\mu / z_1^\mu)^{n_2}, \\
&\tilde\calH_{34}^\text{IR} = -z_1^\alpha z_2^{-\mu} \sum\limits_{n_i = 0}^\infty \tfrac{\Gamma(-\mu n_2 - \nu n_1 - \alpha)}{\Gamma(-\mu n_2 - \nu n_1)} \tfrac{(-z_1^\nu)^{n_1}}{n_1!} (-z_1^\mu / z_2^\mu)^{n_2}.
\end{aligned}\end{equation}

Next, we have three elementary sectors, shown in different colors in Fig.~\ref{Fig2c}. They correspond to three representations in the form of cluster sums:
\begin{align}
\tilde H_3(\bm{z}) &= \tilde\calH_{14}^\lambda + \tilde\calH_{12}^\tau + \tilde\calH_{34}^\text{IR} \nonumber \\
&= \tilde\calH_{24}^{\lambda+\text{IR}} + \tilde\calH_{12}^\tau = \tilde\calH_{13}^{\tau + \lambda} + \tilde\calH_{34}^\text{IR}. \label{ComplRepresentations}
\end{align}

The vector $\bm{A}$ \eqref{Param1} in this case does not depend of the parameters $\nu$ and $\mu$, it coincides with the vector $\bm{A}_4$ and lies exactly between sectors I and II. Therefore, the cluster series $\tilde\calH_{12}^\tau$ converges everywhere, the series $\tilde\calH_{13}^{\tau+\lambda}$ is asymptotic, $\tilde\calH_{14}^\lambda$, $\tilde\calH_{34}^\text{IR}$, and $\tilde\calH_{24}^{\lambda+\text{IR}}$ converge in complimentary regions of $\bm{z}$-space.

\begin{figure}
     \centering
     \begin{subfigure}[b]{0.34\textwidth}
         \centering
         \includegraphics[width=\textwidth]{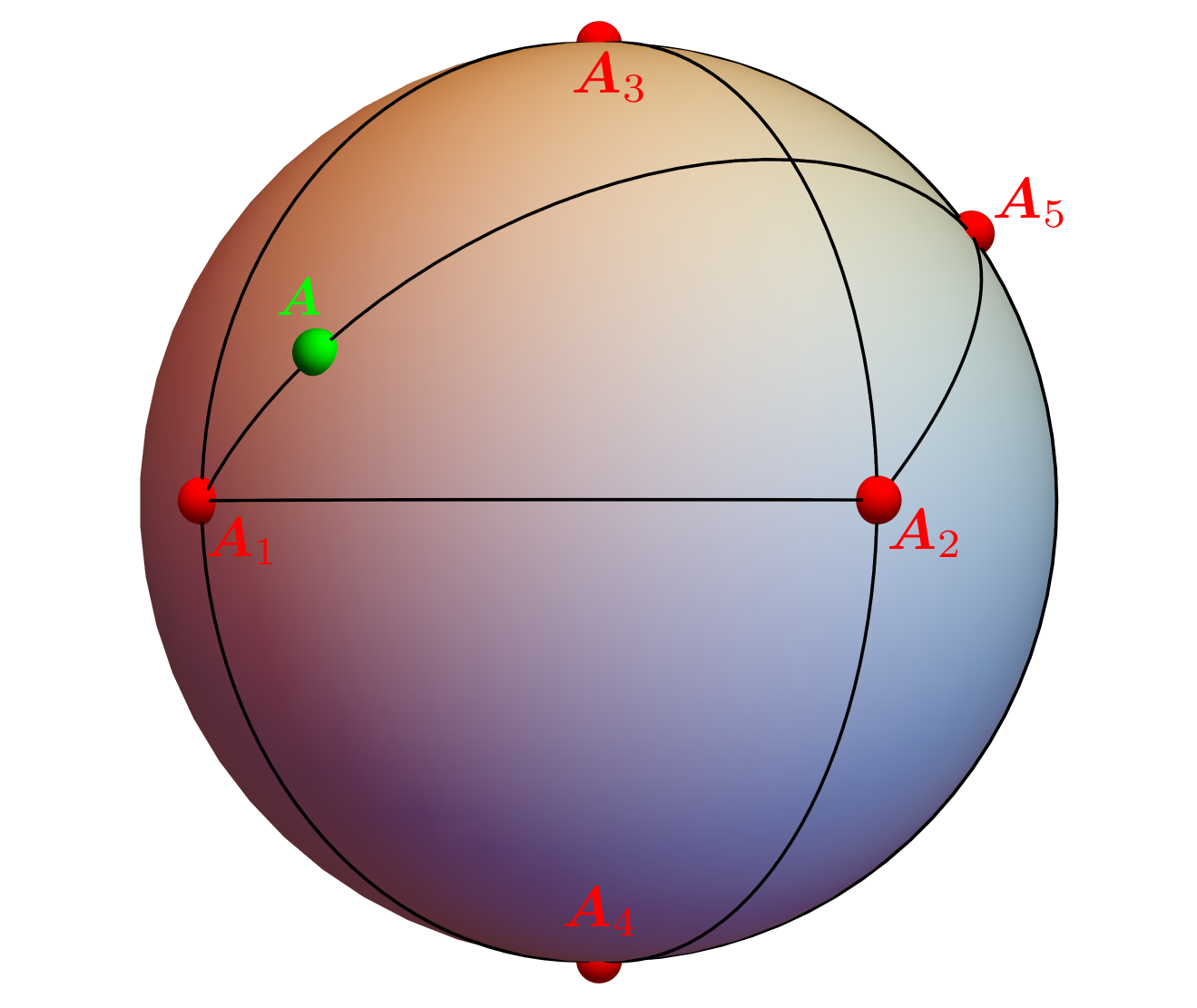}
         \caption{\footnotesize{Vectors $\bm{A}_i$ and $\bm{A}$ normalized to a unit sphere and corresponding elementary sectors.}}
         \label{Fig3a}
     \end{subfigure}
     \hfill
     \begin{subfigure}[b]{0.34\textwidth}
         \centering
         \includegraphics[width=\textwidth]{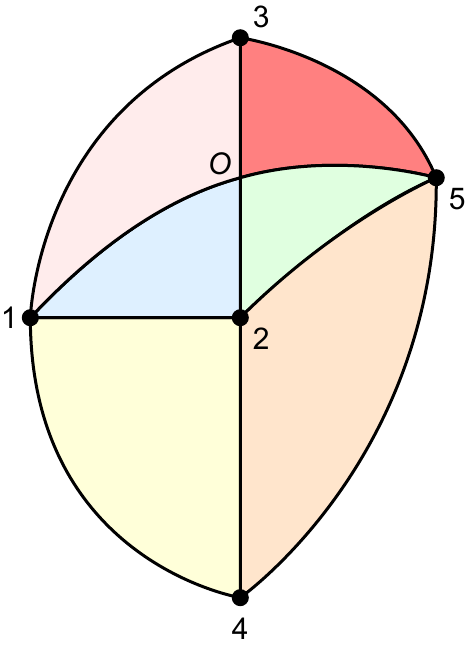}
         \caption{\footnotesize{Minimal intersection regions from Fig. \ref{Fig3a}, projected to a plane for greater clarity.}}
         \label{Fig3b}
     \end{subfigure}
    \caption{\footnotesize{Vectors $\bm{A}_i$ and the corresponding minimal intersection regions for the 3-fold MB integral \eqref{HibridComplete3}.}}
        \label{Fig3}
\end{figure}

\paragraph{Limits for the basis and complementary kernels.}

It is easy to obtain the following expression for the regular kernel:
\begin{align}
&\bbL_\alpha\!\Big[\frac{e^{-\tau F^\nu}}{F^\mu + \lambda}\Big] = \frac{\tau^\frac{\mu-\alpha}{\nu}}{(4\pi)^{d/2} \Gamma(\alpha)}\, \Phi(z), \quad z = \lambda^{1/\mu}\tau^{1/\nu}, \\
&\phi(s) = \frac{1}{\mu\nu} \Gamma\!\left(\frac{s}{\mu}\right)\, \Gamma\!\left(1 - \frac{s}{\mu}\right)\, \Gamma\!\left(\frac{s+\alpha-\mu}{\nu}\right).
\end{align}
Closing the integration contour for $\Phi(z)$ on the right and left we obtain two representations
\begin{equation}
\Phi(z) = \calH^\tau + \calH^\lambda = \calH^{\tau + \lambda},
\end{equation}
where
\begin{equation}\begin{aligned}
&\calH^\tau(z) = \frac{1}{\nu} \sum\limits_{n=0}^\infty \Gamma\!\left(\tfrac{\alpha-\mu(n+1)}{\nu}\right)\, (-z^\mu)^n, \\
&\calH^\lambda(z) = \frac{\pi z^{\alpha-\mu}}{\mu} \sum\limits_{n=0}^\infty \sin^{-1}\!\left(\tfrac{\pi}{\mu}(\alpha+\nu n)\right)\, \frac{(-z^\nu)^n}{n!}, \\
&\calH^{\tau + \lambda}(z) = -\frac{z^{-\mu}}{\nu} \sum\limits_{n=0}^\infty \Gamma\!\left(\tfrac{\alpha+\mu n}{\nu}\right)\, (-z^{-\mu})^n.
\end{aligned}\end{equation}

Calculating the limits of the corresponding contributions $\bbL_\alpha^\tau$ and $\bbL_\alpha^\lambda$ as $\tau, \lambda\to 0$ and comparing the resulting expressions with \eqref{NulLKernellAppEq1} and \eqref{ResolventRegularKernel}, we obtain
\begin{equation} \label{bbLlambdatauLimits} \begin{aligned}
&\bbL_\alpha^\tau\!\Big[\frac{e^{-\tau F^\nu}}{F^\mu + \lambda}\Big] \xrightarrow[\tau,\lambda\to0]{} \bbL_\alpha\!\Big[\frac{e^{-\tau F^\nu}}{F^\mu}\Big], \\
&\bbL_\alpha^\lambda\!\Big[\frac{e^{-\tau F^\nu}}{F^\mu + \lambda}\Big] \xrightarrow[\tau,\lambda\to0]{} \bbL_\alpha\!\Big[\frac{1}{F^\mu+\lambda}\Big].
\end{aligned} \end{equation}

Next, we calculate the limits for $\bbB_\alpha$ and $\bbM_\alpha$ as $\tau, \lambda\to 0$. The three limits for $\bbB_\alpha$ are:
\begin{equation} \label{bbBlambdatauLimits} \begin{aligned}
&\bbB_\alpha^\tau\!\Big[ \frac{e^{-\tau F^\nu}}{F^\mu + \lambda} \Big| \sigma \Big] \xrightarrow[\lambda\to0]{} \bbB_\alpha\!\Big[ \frac{e^{-\tau F^\nu}}{F^\mu} \Big| \sigma \Big], \\
&\bbB_\alpha^\lambda\!\Big[ \frac{e^{-\tau F^\nu}}{F^\mu + \lambda} \Big| \sigma \Big] \xrightarrow[\lambda\to0]{} \bbL_\alpha\!\Big[ \frac{1}{F^\mu+\lambda} \Big], \\
&\bbB_\alpha\!\Big[ \frac{e^{-\tau F^\nu}}{F^\mu + \lambda} \Big| \sigma \Big] \xrightarrow[\tau\to0]{} \bbB_\alpha\!\Big[ \frac{1}{F^\mu+\lambda} \Big| \sigma \Big].
\end{aligned} \end{equation}
In the limit $\lambda\to0$, the cluster series $\calH_{12}^\tau$ and $\calH_{24}^{\overline{\tau}}$ generate two representations via the series $\calH_1^\text{reg}$ and $\calH_2^{\overline{\text{reg}}}$ in the formula \eqref{CalEmu2}, and the leading term of the series $\calH_{14}^\lambda$ leads to the expression \eqref{ResolventRegularKernel}. And in the limit $\tau\to0$, the three series $\calH_{14}^\lambda$, $\calH_{24}^{\overline{\tau}}$, and $\calH_{34}^{\overline{\tau+\lambda}}$ generate the series $\calH_1^\text{reg}$, $\calH_2^\text{UV}$, and $\calH_3^\text{reg+UV}$, respectively, in the formula \eqref{Gexpansions}.

Similarly, for $\bbM_\alpha$ we get the other three limits:
\begin{equation} \label{bbMlambdatauLimits} \begin{aligned}
&\bbM_\alpha\!\Big[ \frac{e^{-\tau F^\nu}}{F^\mu + \lambda} \Big| m^2 \Big] \xrightarrow[\lambda\to0]{} \bbM_\alpha\!\Big[ \frac{e^{-\tau F^\nu}}{F^\mu} \Big| m^2 \Big], \\
&\bbM_\alpha^{\lambda+\text{IR}}\!\Big[ \frac{e^{-\tau F^\nu}}{F^\mu + \lambda} \Big| m^2 \Big] \xrightarrow[\tau\to0]{} \bbM_\alpha\!\Big[ \frac{1}{F^\mu+\lambda} \Big| m^2 \Big], \\
&\bbM_\alpha^\tau\!\Big[ \frac{e^{-\tau F^\nu}}{F^\mu + \lambda} \Big| m^2 \Big] \xrightarrow[\tau\to0]{} \bbL_\alpha\!\Big[ \frac{e^{-\tau F^\nu}}{F^\mu} \Big].
\end{aligned} \end{equation}
In this case, in the limit $\lambda\to0$, the cluster series $\tilde\calH_{12}^\tau$ and $\tilde\calH_{24}^{\lambda+\text{IR}}$ give the series $\tilde\calH_1^\text{reg}$ and $\tilde\calH_2^\text{IR}$ in the formula \eqref{tildeEexp}. In the limit $\tau\to0$, the three series $\tilde\calH_{14}^\lambda$, $\tilde\calH_{34}^\text{IR}$, and $\tilde\calH_{24}^{\lambda+\text{IR}}$ generate the series $\tilde\calH_1^\text{reg}$, $\tilde\calH_2^\text{IR}$, and $\tilde\calH_3^\text{reg+IR}$, respectively, in the formula \eqref{Gexpansions}, and the leading term of the series $\tilde\calH_{12}^\tau$ leads to the expression \eqref{NulLKernellAppEq1}.

In the formulas \eqref{bbLlambdatauLimits}-\eqref{bbMlambdatauLimits}, the $\tau$-contributions dominate in the relevant region $\Re(\alpha-\mu)>0$, and the $\lambda$-contributions dominate in the irrelevant region $\Re(\alpha-\mu)<0$.

\paragraph{The complete massive kernel} is the only example of a 3-fold MB integral that we consider in this paper. It can be written as follows:
\begin{align}
&\bbW_\alpha\!\Big[ \frac{e^{-\tau F^\nu}}{F^\mu + \lambda} \Big| \sigma, m^2 \Big] =  \frac{\tau^\frac{\mu-\alpha}{\nu}}{(4\pi)^{d/2}}\; H_4(\bm{z}), \label{HibridComplete1} \\
&\bm{z} = \left(\frac{\sigma}{2\tau^{1/\nu}}, m^2\tau^{1/\n}, \lambda^{1/\m} \tau^{1/\nu}\right), \\
&h_4(\bm{s}) = \Gamma(s_1) \Gamma(s_2)\Gamma\!\left(\tfrac{s_3}{\mu}\right) \Gamma\!\left(1-\tfrac{s_3}{\mu}\right) \frac{\Gamma\!\left(\tfrac{s_2+s_3-s_1+\alpha-\mu}{\nu}\right)}{\mu\nu\,\Gamma(s_2 - s_1+\alpha)}. \label{HibridComplete3}
\end{align}

The function $h_4(\bm{s})$ has the following parameters:
\begin{equation} \label{Amatrix4} \begin{aligned}
\bm{\alpha} &= (1, 1, 1, 1, 1, -1), \\
\bm{a} &= \left(0, 0, 0, -1, \tfrac{\mu-\alpha}{\nu}, -\alpha\right), \\
A &= \begin{pmatrix}
1 & 0 & 0 & 0 & -1/\nu & -1 \\
0 & 1 & 0 & 0 & 1/\nu & 1\\
0 & 0 & 1/\mu & -1/\m & 1/\n & 0\\
\end{pmatrix}.
\end{aligned}\end{equation}

The vectors $\bm{A}_i$ are shown in Fig.~\ref{Fig3}. It can be seen that we have 7 clusters:
\begin{equation} \label{Clusters}
\{123\}, \{124\}, \{125\}, \{135\}, \{145\}, \{235\}, \{245\}.
\end{equation}
Instead of ``elementary sectors'' like in the 2-fold case, in the 3-fold case we must consider the minimal intersection regions of the corresponding spherical triangles. Fortunately, in the case under consideration, these regions will also be spherical triangles (although in the general case this is obviously not necessarily so), which we will also denote by their three angles. It is easy to see that in our case there are 6 such minimal intersection regions. Let us list them, introducing an additional point $O$ at the intersection of the arcs $\{15\}$ and $\{23\}$:
\begin{equation} \label{Intersections}
\{O12\}, \{O13\}, \{O25\}, \{O35\}, \{124\}, \{245\}.
\end{equation}
Each triangle from \eqref{Intersections} generates its own representation of the function $H_4(\bm{z})$ as a sum over the triangles from \eqref{Clusters} that contain it. We have:
\begin{align}
H_4(\bm{z}) &= \calH_{135}^{\lambda+\text{IR}} + \calH_{235}^{\overline{\tau}} = \calH_{125}^\lambda + \calH_{235}^{\overline{\tau}} + \calH_{145}^\text{IR} \nonumber \\
&= \calH_{135}^{\lambda+\text{IR}} + \calH_{123}^\tau = \calH_{125}^\lambda + \calH_{123}^\tau + \calH_{145}^\text{IR} \nonumber \\
&= \calH_{124}^{\tau+\lambda} + \calH_{145}^\text{IR} = \calH_{245}^{\overline{\tau + \lambda}} + \calH_{145}^\text{IR}. \label{AllRepresentations}
\end{align}

As before, it is easy to find families of simple poles generated by each of the clusters:
\begin{equation}\begin{aligned}
&\bm{s}_{123}(\bm{n}) = \big(-n_1, -n_2, -\mu n_3\big), \\
&\bm{s}_{124}(\bm{n}) = \big(-n_1, -n_2, \mu(n_3+1)\big), \\
&\bm{s}_{125}(\bm{n}) = \big(-n_1, -n_2, \mu-\alpha-n_1+n_2-\nu n_3\big), \\
&\bm{s}_{135}(\bm{n}) = \big(-n_1, -\alpha-n_1-\nu n_2+\mu(n_3+1), -\mu n_3\big), \\
&\bm{s}_{145}(\bm{n}) = \big(-n_1, -\alpha-n_1-\nu n_2-\mu n_3, \mu(n_3+1)\big), \\
&\bm{s}_{235}(\bm{n}) = \big(\alpha+\nu n_1-n_2-\mu(n_3+1), -n_2, -\mu n_3\big), \\
&\bm{s}_{245}(\bm{n}) = \big(\alpha+\nu n_1-n_2+\mu n_3, -n_2, \mu(n_3+1)\big),
\end{aligned}\end{equation}
and calculate the corresponding cluster series:
\begin{widetext}
\begin{equation} \label{AllSeries}\begin{aligned}
&\calH_{123}^\tau(\bm{z}) = \sum\limits_{n_i=0}^\infty \frac{\Gamma\!\left(\tfrac{n_1-n_2-\mu(n_3+1)+\alpha}{\nu}\right)}{\nu\Gamma(n_1-n_2+\alpha)} \frac{(-z_1)^{n_1}}{n_1!} \frac{(-z_2)^{n_2}}{n_2!} (-z_3^\mu)^{n_3}, \\
&\calH_{124}^{\tau+\lambda}(\bm{z}) = -z_3^{-\mu} \sum\limits_{n_i=0}^\infty \frac{\Gamma\!\left(\tfrac{n_1-n_2+\mu n_3+\alpha}{\nu}\right)}{\nu\Gamma(n_1-n_2+\alpha)} \frac{(-z_1)^{n_1}}{n_1!} \frac{(-z_2)^{n_2}}{n_2!} (-z_3^{-\mu})^{n_3}, \\
&\calH_{125}^\lambda(\bm{z}) = z_3^{\alpha-\mu} \sum\limits_{n_i=0}^\infty \frac{\pi \Gamma^{-1}(n_1-n_2+\alpha) /\mu}{\sin\tfrac{\pi}{\mu}(\alpha+n_1-n_2+\nu n_3)} \frac{(-z_1z_3)^{n_1}}{n_1!} \frac{(-z_2/z_3)^{n_2}}{n_2!} \frac{(-z_3^\nu)^{n_3}}{n_3!}, \\
&\calH_{135}^{\lambda+\text{IR}}(\bm{z}) = z_2^{\alpha-\mu} \sum\limits_{n_i=0}^\infty \frac{\Gamma\big(-\alpha-n_1-\nu n_2+\mu(n_3+1)\big)}{\Gamma\big(-\nu n_2+\mu(n_3+1)\big)} \frac{(-z_1z_2)^{n_1}}{n_1!} \frac{(-z_2^\nu)^{n_2}}{n_2!} (-z_3^\mu/z_2^\mu)^{n_3}, \\
&\calH_{145}^\text{IR}(\bm{z}) = -z_2^\alpha z_3^{-\mu} \sum\limits_{n_i=0}^\infty \frac{\Gamma(-\alpha-n_1-\nu n_2-\mu n_3)}{\Gamma(-\nu n_2-\mu n_3)} \frac{(-z_1z_2)^{n_1}}{n_1!} \frac{(-z_2^\nu)^{n_2}}{n_2!} (-z_2^\mu/z_3^\mu)^{n_3}, \\
&\calH_{235}^{\overline{\tau}}(\bm{z}) = -z_1^{\mu-\alpha} \sum\limits_{n_i=0}^\infty \frac{\Gamma\big(\alpha+\nu n_1-n_2-\mu(n_3+1)\big)}{\Gamma\big(-\nu n_1+\mu(n_3+1)\big)} \frac{(-z_1^{-\nu})^{n_1}}{n_1!} \frac{(-z_1z_2)^{n_2}}{n_2!} (-z_1^\mu z_3^\mu)^{n_3}, \\
&\calH_{245}^{\overline{\tau + \lambda}}(\bm{z}) = z_1^{-\alpha}z_3^{-\mu} \sum\limits_{n_i=0}^\infty \frac{\Gamma(\alpha+\nu n_1-n_2+\mu n_3)}{\Gamma(-\nu n_1-\mu n_3)} \frac{(-z_1^{-\nu})^{n_1}}{n_1!} \frac{(-z_1z_2)^{n_2}}{n_2!} (-z_1^{-\mu} z_3^{-\mu})^{n_3}.
\end{aligned}\end{equation}
\end{widetext}

The vector $\bm{A}$ \eqref{Param1} takes the form
\begin{equation}
\bm{A} = \begin{pmatrix} 2-1/\n\\
1/\n\\
1/\n\end{pmatrix}.
\end{equation}
If we normalize it to the unit sphere, then as the parameter $\nu>0$ changes, it will move along the arc $\{15\}$. Consequently, the cluster series $\calH_{124}^{\tau+\lambda}$ and $\calH_{245}^{\overline{\tau + \lambda}}$ are always asymptotic, and the series $\calH_{125}^\lambda$, $\calH_{135}^{\lambda+\text{IR}}$ and $\calH_{145}^\text{IR}$ converge, each in its own region. For $0<\nu<1/2$, the vector $\bm{A}$ lies on the arc $\{01\}$. Accordingly, the series $\calH_{235}^{\overline{\tau}}$ converges everywhere, and the series $\calH_{123}^\tau$ is asymptotic. For $1/2<\nu<\infty$, the vector $\bm{A}$ lies on the arc $\{05\}$, and therefore the situation is reversed: $\calH_{235}^{\overline{\tau}}$ becomes asymptotic, and $\calH_{123}^\tau$ begins to converge everywhere.

\paragraph{Limits for the complete massive kernel.}

Finally, the limits as $\tau, \lambda\to0$ have a very simple form
\begin{equation} \begin{aligned}
&\bbW_\alpha\!\Big[ \frac{e^{-\tau F^\nu}}{F^\mu + \lambda} \Big| \sigma, m^2 \Big] \xrightarrow[\sigma\to0]{} \bbM_\alpha\!\Big[ \frac{e^{-\tau F^\nu}}{F^\mu + \lambda} \Big| m^2 \Big], \\
&\bbW_\alpha^\text{reg}\!\Big[ \frac{e^{-\tau F^\nu}}{F^\mu + \lambda} \Big| \sigma, m^2 \Big] \xrightarrow[m^2\to0]{} \bbB_\alpha\!\Big[ \frac{e^{-\tau F^\nu}}{F^\mu + \lambda} \Big| \sigma \Big], \\
&\bbW_\alpha\!\Big[ \frac{e^{-\tau F^\nu}}{F^\mu + \lambda} \Big| \sigma, m^2 \Big] \xrightarrow[\tau\to0]{} \bbW_\alpha\!\Big[ \frac{1}{F^\mu + \lambda} \Big| \sigma, m^2 \Big], \\
&\bbW_\alpha\!\Big[ \frac{e^{-\tau F^\nu}}{F^\mu + \lambda} \Big| \sigma, m^2 \Big] \xrightarrow[\lambda\to0]{} \bbW_\alpha\!\Big[ \frac{e^{-\tau F^\nu}}{F^\mu} \Big| \sigma, m^2 \Big].
\end{aligned} \end{equation}

Indeed, in the limit of $\sigma\to0$, we have obvious correspondences between series with matching superscripts in the formulas \eqref{AllSeries} and \eqref{ClSeries2}. Then, the three representations in the formula \eqref{AllRepresentations}, corresponding to the minimal intersection regions $\{O12\}$, $\{O13\}$, and $\{124\}$, generate three representations in the formula \eqref{ComplRepresentations}.

A similar correspondence between series with matching superscripts in the formulas \eqref{AllSeries} and \eqref{ClSeries1} holds in the massless limit $m^2\to0$. Then the regular contributions in the four representations in \eqref{AllRepresentations} corresponding to the minimal intersection regions $\{O12\}$, $\{O25\}$, $\{124\}$, and $\{245\}$ generate four representations in the formula \eqref{BasisRepresentations}.

In the limit $\lambda\to0$, we have the following correspondence between the series in the formulas \eqref{AllSeries} and \eqref{ClusterH}
\begin{equation}
\calH_{123}^\tau\to \calH_{12}^\text{reg}, \quad \calH_{235}^{\overline{\tau}} \to \calH_{23}^{\overline{\text{reg}}}, \quad \calH_{135}^{\lambda+\text{IR}} \to \calH_{13}^\text{IR}.
\end{equation}
And the two representations in the formula \eqref{AllRepresentations}, corresponding to the minimal intersection regions $\{O13\}$ and $\{O35\}$, generate exactly the two representations in the formula \eqref{ClusterH}.

Finally, in the limit $\tau\to0$, we have the following correspondence between the series in the formulas \eqref{AllSeries} and \eqref{ResolvSeries}
\begin{gather}
\calH_{125}^\lambda \to \calH_{12}^\text{reg}, \quad \calH_{145}^\text{IR} \to \calH_{13}^\text{IR}, \quad \calH_{235}^{\overline{\tau}} \to \calH_{24}^\text{UV}, \nonumber \\
\calH_{135}^{\lambda+\text{IR}} \to \calH_{14}^\text{reg+IR}, \quad \calH_{245}^{\overline{\tau+\lambda}} \to \calH_{23}^\text{reg+UV}.
\end{gather}
Then the three different representations in the formula \eqref{AllRepresentations}, corresponding to the three minimal intersection regions $\{O35\}$, $\{O25\}$, and $\{245\}$, generate exactly the three representations in the formula \eqref{GclusterSums}.

\section{Resonant case example} \label{ResonantSec}

In this section we explicitly compute series representations for MB integrals in the resonant case which arises when several simple poles coalesce for specific values of parameters $\bm{a}$ in~\eqref{GammaProduct}-\eqref{NfoldIntDef}. For the basis and complete kernels' MB representations we have considered, this corresponds to integer values of parameters $\m$, $\n$, and $\a$, so in this section we will assume $\m,\n,\a\in\bbZ$ unless it is explicitly stated otherwise. As we have noted in Sec.~\ref{SeriesSubSec}, the resonant case is of particular significance from the physical point of view, since, for example, parameter $\alpha$ is defined as $\alpha=d/2-n$, where $n\in\bbZ_{\geq0}$ corresponds to the ordinal number of the HaMiDeW coefficient, so, naturally, it must take integer values. However, as we showed in~\cite{BKW25a}, in many cases basis kernels contain IR divergences, which in the MB representation manifest themselves as contour pinches. As these pinches only arise in the resonant case, the non-resonant expressions serve as regularized versions of corresponding resonant series. For example, representations with $\alpha\notin\bbZ$ can be viewed as the dimensionally regularized versions of the corresponding IR-divergent resonant series. Consistency of this procedure with massive regularization was directly verified in~\cite{BKW25a}, by comparing divergent and finite terms of resonant series for IR-divergent basis kernels $\bbB_{\alpha+\epsilon}[f|\sigma]$, $\epsilon\to0$, with massless limit $m^2\to0$ of complete kernel resonant series $\bbW_{\alpha}[f|\sigma,m^2]$ for several examples of operator functions $f$. These explicit checks relied on series representations of certain double MB integrals in the resonant case, which were stated without a proper derivation. The goal of this section is to fill this gap.

An important thing to note is that strictly speaking, there is no need to explicitly obtain resonant series representations separately, since due to the principle of continuity for local residue~\cite{GriffithsHarris} they can be obtained from the corresponding non-resonant series representations by taking the appropriate limit in the parameter space.  Therefore, in some sense, the non-resonant case is more fundamental, even though it does not directly correspond to any physical situation. Despite this fact we choose to do a separate resonant computation anyway for reasons of completeness and additional verification. Moreover, in the computation below we make use of the multidimensional residue, so this section might be useful to some readers who would like to familiarize themselves with this object. 

The only series representation of resonant $N$-fold MB integral with $N>1$ we used in~\cite{BKW25a} is one for the complete kernel $\bbW_\alpha[f|\sigma, m^2]$ for the operator function $f(\h F)=e^{-\tau\hat F}/\hat F$. Its MB representation is given by \eqref{tildeE1}-\eqref{tildeE3}, which for $\m=\n=1$ reads
\begin{align}
    \bbW_{\a}\lb\frac{e^{-\tau F}}{F}\Bigg| \sigma,m^2 \rb &=\frac{\tau^{1-\alpha}}{(4\pi)^{d/2}}\int\limits_{C}\frac{ds_1ds_2}{(2\pi i)^2}z_1^{-s_1}z_2^{-s_2} \nonumber \\
&\times\frac{\Gamma(s_1)\Gamma(s_2-\a)}{s_2-s_1-1}, \label{h3mu1nu1}
\end{align}
where $\lp z_1,z_2\rp=\lp\sigma/2\tau,m^2\tau\rp$. An important observation one can make is that for all integer values of $\alpha$ the non-splitting 2-cycle $C$ in the integral~\eqref{h3mu1nu1} can be chosen straight, so, in accordance with the general theory outlined in Sec.~\ref{DefSubSec}, no 2-cycle deformation is necessary and one can simply choose $C=\bm{\g}+i\bbR^2$ with $\bm\gamma\in\bbR^2$. This somewhat simplifies the series evaluation procedure in the resonant case, which we briefly outline below.

\subsection{Calculation scheme}

Consider a double Mellin-Barnes integral with a straight contour
\begin{equation} \label{tempint}
        H(z_1,z_2)\!=\int\limits_{\bm{\gamma}+i\bbR^2}\!\!\frac{d^2s}{(2\pi i)^2}\,h(\bm{\alpha},A,\bm{a}|s_1,s_2)\,\bm{z}^{-\bm s}
\end{equation}
where $\bm z=(z_1,z_2)$ and $\bm s=(s_1,s_2)$ lie in $\bbC^2$ and $h(\bm{\alpha},A,\bm{a}|s_1,s_2)$ is as in~\eqref{GammaProduct}, however, unlike in previous sections, parameters $\bm{a}$ can be arbitrary, so that more than two polar hyperplanes $L_{i,n}$~\eqref{Hyperplanes} can intersect. The point $\bm{\gamma}\in \bbR^2$ is chosen as before: so that it lies to one side of every $L_{i,n}$. The difference with the non-resonant case appears at the step we are calculating residues at intersections of hyperplanes, which we will cover shortly, but first, it is convenient to move the singularities of the integrand to the origin. Therefore, the method for evaluating resonant two-fold MB integrals with a straight contour can be formulated as follows~\cite{Zhdanov1998}.

First, perform the change of variables to move the singularities of the integrand~\eqref{tempint} $\bm s^*=\bm s^*(\bm n)$, $\bm n\in \bbZ^2_{\geq0}$ to the origin $0$ for a fixed $\bm n$. Then pick out these singularities using the reflection formula~(\ref{ReflectionFormula}) for the gamma function,
\begin{equation}
    \Gamma(s-k)=\frac{(-1)^k}{s}\frac{\Gamma(1-s)\Gamma(1+s)}{\Gamma(k+1-s)},\qquad\forall k\in\bbZ.
\end{equation}
After this, the integrand will have the singularities explicitly extracted to the denominator. In the non-resonant case it would have the simplest form $w(s_1, s_2)/s_1^ns_2^m$ with $w(s_1, s_2)$ holomorphic at the origin and $n,m$ being some integers. In that case the residue at $\bm s=0$ can be understood as the iterated residue [see~\eqref{ClusterResidues}]. However, in our case the denominator can look more complicated, containing more than two factors, for example, one could have the integrand as
\begin{equation} \label{res_pqmn}
    \frac{w(s_1,s_2)}{(s_1+a s_2)^p(s_1-b s_2)^q s_1^n s_2^m},
\end{equation}
where $w(s_1,s_2)$ is holomorphic at the origin and $p,q,m,n$ are integers. Presence of these additional factors in the denominator makes it impossible to interpret the multivariate residue at this point simply as an iterated 1-dimensional residue and it should be understood as the local Grothendieck residue of the corresponding holomorphic 2-form $\omega(\bm s)$~\cite{GriffithsHarris}. For $p=q=0$ it reduces to the iterated residue,
\begin{align}
    \Res_{\bm{s}=0} \,\omega (s_1,s_2) &= \Res_{\bm s=0}\frac{w(s_1,s_2)}{s_1^n s_2^m}ds_1\wedge ds_2 \nonumber \\
    &=\frac{\pp_{s_1}^{n-1}\pp_{s_2}^{m-1}w(s_1,s_2)}{(n-1)!(m-1)!}\Big\vert_{\bm s=0}.
\label{res_triv}
\end{align}
{However, in the general case~(\ref{res_pqmn}) one needs to make use of the so-called \textit{transformation formula}, which may be formulated as follows~\cite{Tsikh1992, GriffithsHarris, Larsen2018}:

Suppose that two $\bbC^N\to\bbC^N$ mappings $\bm{f}(\bm s) = \big(f_i(\bm s)\big)$ and $\bm{g}(\bm{s}) = \big(g_i(\bm s)\big)$, $i = 1, \ldots, N$, both have common isolated zeros at the origin $\bm{s} = 0 \in\bbC^N$. Also let $g_i(\bm s)=T_i^j(\bm s) f_j(\bm s)$, where the holomorphic $(N\times N)$ transformation matrix $T_i^j(\bm{s})$ can be degenerate at the origin. Then for any holomorphic $w(\bm{s})$ one has
\begin{align}
&\Res_{\bm{s}=0}\frac{w(\bm{s})}{f_1(\bm{s})\cdots f_N(\bm{s})}\, ds_1\wedge\cdots\wedge ds_N \nonumber \\
= &\Res_{\bm{s}=0}\frac{\det T_i^j(\bm{s})\; w(\bm{s})}{g_1(\bm{s})\cdots g_N(\bm{s})}\, ds_1\wedge\cdots\wedge ds_N.
\label{transformation_law}
\end{align}

This is essentially an $N$-dimensional generalization of a trivial 1-dimensional fact:
\begin{equation}
    \Res_{s=0}\frac{w(s)}{f(s)}\,ds=\Res_{s=0}\frac{w(s)}{g(s)}T(s)\,ds,\quad T(s)=\frac{g(s)}{f(s)}.
\end{equation}

The sketch of a proof is as follows: first assume that both $\det T_i^j$ and $\det \partial f_i/\partial s_j$ are nonzero at $\bm{s}=0$. Then one has
\begin{align}
    \det\frac{\partial g_i}{\partial s_k}\Big|_{\bm{s}=0}
&= \det\Big[\frac{\partial T_i^j}{\partial s_k}f_j\Big|_{\bm{s}=0} +T_i^j\frac{\partial f_{j}}{\partial s_k}\Big|_{\bm{s}=0}\Big] \nonumber \\
    &=\det T_i^j(0) \; \det\frac{\partial f_j}{\partial s_k}\Big|_{\bm{s}=0}. \label{f_g_jacobians}
\end{align}

Moreover, for nonzero Jacobian through a simple change of variables $\bm s\mapsto \bm t=\bm f(\bm s)$ under the integral sign, we have
\begin{align}
    &\Res_{\bm{s}=0}\frac{w(\bm s)}{f_1(\bm s)\dots f_N(\bm s)}ds_1\wedge\dots\wedge ds_N \nonumber \\
    &=\frac{1}{(2\pi i)^N}\int \frac{w(\bm s)}{f_1(\bm s)\dots f_N(\bm s)}{ds_1\wedge\dots \wedge ds_N} \nonumber \\
    &=\frac{1}{(2\pi i)^N}\int \frac{w\big(\bm f^{-1}(\bm t)\big)}{\det({\partial  t_i}/{\partial f^{-1}_j})}\frac{dt_1\wedge\dots \wedge dt_N}{t_1\dots t_N} \nonumber \\
    &=\frac{w\big(\bm f^{-1}(\bm t)\big)}{\det({\partial  t_i}/{\partial f^{-1}_j})}\Bigg|_{\bm t=0}=\frac{w(\bm s)}{\det (\partial f_i/\partial s_j)}\Bigg\vert_{\bm{s}=0}.
\end{align}
Similarly, for the r.h.s. of~(\ref{transformation_law}) we have:
\begin{align}
    &\Res_{\bm{s}=0}\frac{\det T_i^j(\bm{s}) \; w(\bm{s})}{g_1(\bm{s})\dots g_N(\bm{s})}\,ds_1\wedge\dots\wedge ds_N \nonumber \\
    &=\frac{\det T_i^j(\bm{s}) \; w(\bm s)}{\det (\partial g_i/\partial s_j)}\Bigg\vert_{\bm{s}=0}=\frac{w(\bm s)}{\det (\partial f_i/\partial s_j)}\Bigg\vert_{\bm{s}=0},
\end{align}
with the last equality holding due to~(\ref{f_g_jacobians}). This completes the proof of the transformation law in the case $\det T_i^j(0)\neq0$ and $\det \partial f_i/\partial s_j\big|_{\bm{s}=0}\neq0$. The general case is a bit more subtle and can be proven by reduction to the nondegenerate case considered above. The idea is as follows: first, for a zero Jacobian and invertible $T_i^j(\bm s)$ one deforms the maps $\bm f(\bm s)$ slightly so that the Jacobian is nonzero and by continuity obtains the needed result. Finally for a degenerate $T_i^j(\bm s)$ one considers a continuous family $T_i^j(\bm s|\varepsilon)$ such that $\det T_i^j(\bm s|\varepsilon)\neq0$ for $\varepsilon\neq0$, and, after showing that residues vanish in all zeroes of $g_i(\bm s|\varepsilon)\equiv T_i^j(\bm s|\varepsilon) f_j(\bm s)$ except for $\bm{s}=0$ one takes the limit $\varepsilon\to0$, finally proving~\eqref{transformation_law}.

Using the transformation law~\eqref{transformation_law} we can essentially reduce a more complicated case like~\eqref{res_pqmn} to a familiar one with $p=q=0$~\eqref{res_triv}. There is, however, another obstacle.}

Note, that in general one could have a product of more than $N$ factors in the denominator (more than two in our 2-fold case), so the choice of \emph{divisors} $\bm{f}(\bm{s})=(f_1(\bm s),f_2(\bm s))$ is ambiguous and will affect the result\footnote{See \cite{Larsen2018} for a topological explanation of this fact.}. In fact, one has to take the integration contour into consideration\footnote{This is due to the fact that the divisor system needs to be compatible with the polyhedron over the boundary of which $\omega$ is being integrated~\cite{Passare1994}.} to determine which of the factors in the denominator of~(\ref{res_pqmn}) constitute $f_1(s_1,s_2)$ and which constitute $f_2(s_1,s_2)$. The rule of thumb is as follows~\cite{Friot2012}: 

\emph{First calculate vector $\bm{A}$ defined in~\eqref{Param1}, then draw a line
\begin{equation}
\ell_{\bm{A}}=\lc A_1\Re s_1+A_2\Re s_2=A_1\g_1+A_2\g_2\rc
\end{equation}
 through the point $\bm{\gamma}$ in the $\Im s_1=\Im s_2=0$ plane and mark a positive direction $\bm{l}_{+}$ on $\ell_{\bm{A}}$ such that the pair $\bm{l}_+$ and $\bm{A}$ gives the same orientation of the plane as do the coordinate axes $\Re s_1$, $\Re s_2$. Then polar lines which intersect the line $\ell_{\bm{A}}$ before the point $\bm{\gamma}$ correspond to $f_1$, while those that intersect $\ell_{\bm{A}}$ after $\bm{\gamma}$ constitute $f_2$. If all polar lines intersect $\ell_{\bm{A}}$ to one side of $\bm{\gamma}$, the corresponding poles are called spurious and do not contribute. Finally, the sum of these residues over $\bm{n}$ yields the sought series representation of the resonant two-fold MB integral.}

For $N=2$, this recipe applies equally to the general nonresonant case \eqref{res_triv} as well as to the degenerate resonant case. In the given formulation, it may seem rather difficult, but its meaning is easily understood by the simplest example, which we will analyze in the next Subsection.

\subsection{Resonant series for the complete kernel of $\exp(-\tau \h F)/\h F$}

We now follow the procedure described in the preceding subsection to find the series representation for the integral~\eqref{h3mu1nu1}. The corresponding holomorphic 2-form reads
\begin{equation}
\omega(s_1,s_2)=z_1^{-s_1}z_2^{-s_2}\frac{\Gamma(s_1)\Gamma(s_2-\a)}{s_2-s_1-1}\,ds_1\wedge ds_2.
\label{omega_form}
\end{equation}
As we have noted above, the non-splitting integration 2-cycle can be chosen to be straight since there exists $\bm{\g}$ such that the arguments of gamma functions in~\eqref{omega_form} are positive. A possible choice of $\bm{\gamma}=(\gamma_1,\gamma_2)$ is $\gamma_1=1/2$, $\gamma_2=\mathrm{max}\lc\a,\m+1/2\rc+1/4=\mathrm{max}\lc \a+1/4,7/4 \rc$. This allows us to implement the technique described in the previous subsection without any modifications.

\begin{figure}
     \centering
    \begin{subfigure}[b]{0.35\textwidth}
         \centering
         \includegraphics[width=\textwidth]{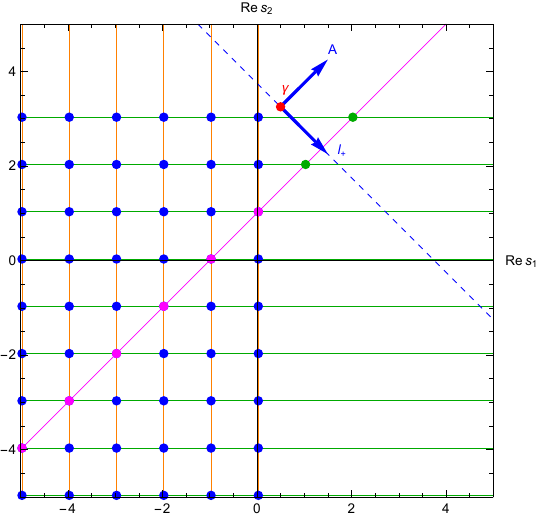}
         \caption{\footnotesize{$\a=3$}}
         \label{fig:divisor_a3}
     \end{subfigure}
     \hfill
     \begin{subfigure}[b]{0.35\textwidth}
         \centering
         \includegraphics[width=\textwidth]{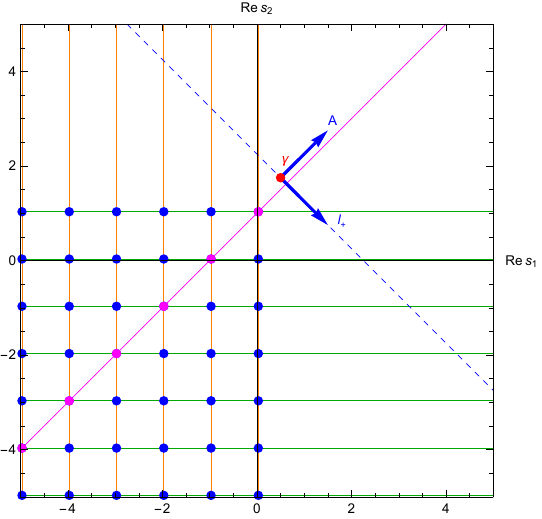}
         \caption{\footnotesize{$\a=1$}}
         \label{fig:divisor_a1}
     \end{subfigure}
     \hfill
     \begin{subfigure}[b]{0.35\textwidth}
         \centering
         \includegraphics[width=\textwidth]{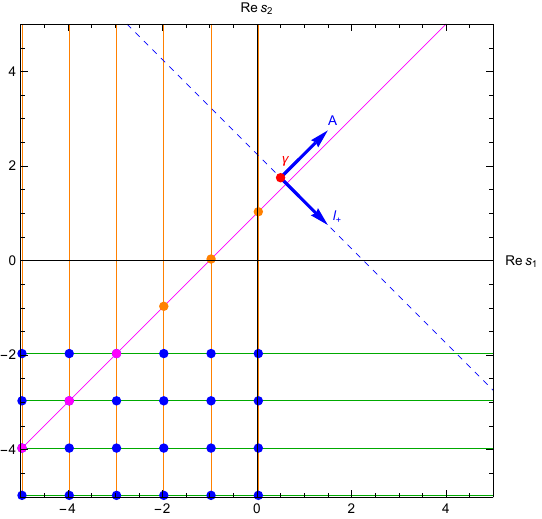}
         \caption{\footnotesize{$\a=-2$}}
         \label{fig:divisor_aneg2}
     \end{subfigure}

    \caption{\footnotesize{Singular structure of the integrand of~(\ref{h3mu1nu1}) in three cases: $\a<1$~(\ref{fig:divisor_aneg2}), $\a>1$~(\ref{fig:divisor_a3}), and $\a=1$~(\ref{fig:divisor_a1}). The orange vertical lines correspond to the singularities of $\Gamma(s_1)$, green lines are the singularities of $\Gamma(s_2-\a)$ and the magenta line correspond to the factor $(s_2-s_1-1)^{-1}$. The point $\bm{\gamma}$ is shown in red, vector $\bm{A}$ is in blue, and the dashed blue line is $\ell_{\bm{A}}$ with its positive direction $\bm{l}_+$ shown also in blue.}}
        \label{fig:divisor_plot}
\end{figure}

Calculating the integral~\eqref{tempint} boils down to finding the sum of local residues of $\omega(s_1,s_2)$ in its poles which are shown as small circles on the plots of singular structure of~\eqref{omega_form} given in Fig.~\ref{fig:divisor_plot} for different integer values of $\alpha$. Theses plots also contain auxiliary constructions: the point $\bm{\gamma}$ (shown in red) and the line $\ell_{\bm{A}}$ ($\bm{A}=(1,1)^T$ in our case) with its orientation vector $\bm{l}_+$.

As can be seen from the Fig.~\ref{fig:divisor_plot}, there are poles of only four different types, which are marked with different colors: blue, magenta, orange, and green. It makes sense to consider these four types separately.

\paragraph{``Blue'' poles.}

First of all, for any value of the parameter $\alpha$, we have an infinite two-parameter family of ``blue'' simple poles formed by the intersection of the polar lines of the gamma functions $\Gamma(s_1)$ and $\Gamma(s_2-\a)$
\begin{equation}
(-n_1, \alpha-n_2), \quad n_{1,2}\in\bbZ_{\geq0},  \quad n_1-n_2 \neq 1-\alpha.
\end{equation}
The residues in them are calculated in the same way as in the non-resonant case and give the following contribution
\begin{equation} \label{BlueSeries}
H_\text{blue} = \sum_{\tln{n_1,n_2=0}{n_1-n_2+\a\neq1}}^{\infty}\frac{z_2^{-\a}}{\a+n_1-n_2-1}\frac{(-z_1)^{n_1}}{n_1!}\frac{(-z_2)^{n_2}}{n_2!}.
\end{equation}

\paragraph{``Orange'' and ``green'' poles.}

Further, for $\alpha\in\bbZ_{<1}$, $(1-\alpha)$ simple ``orange'' poles appear, formed by the intersection of the singular line $s_2 - s_1 = 1$ with the polar lines of the gamma function $\Gamma(s_1)$. The finite contribution from these poles is
\begin{equation} \label{OrangePol}
H_\text{orange} = \frac{1}{z_2} \sum_{n=0}^{-\a}\frac{(-\a-n)!}{n!}(-z_1z_2)^n.
\end{equation}

Conversely, for $\alpha\in\bbZ_{>1}$, $(\alpha-1)$ simple ``green'' poles appear, formed by the intersection of the singular line $s_2 - s_1 = 1$ with the polar lines of the gamma function $\Gamma(s_2-\a)$. However, these polar lines intersect the line $\ell_{\bm A}$ on the same side as the singular line, so, according to the computational scheme outlined in the previous subsection, these poles are ``spurious'' and do not contribute to the answer
\begin{equation}
H_\text{green} = 0.
\end{equation}

\paragraph{``Magenta'' poles for $\alpha\ge1$.}

The three types of poles considered so far are simple, and therefore the residues in them can be understood as iterated, and their calculation is no different from the non-resonant case. The general theory of local Grothendieck residues, necessary for the resonant case, is required for degenerate poles at which more than $N$ polar hyperplanes intersect. In the case under consideration, these will be an infinite one-parameter family of magenta-colored poles, in which, for integer $\alpha$, the singular line $s_2 - s_1 = 1$ and the polar lines of the gamma functions $\Gamma(s_1)$ and $\Gamma(s_2-\a)$ intersect simultaneously.

For $\alpha\ge1$, they are located at the points $(-n,1-n)$, $n\in\bbZ_{\geq0}$. In the variables $t_1=s_1+n$, $t_2=s_2+n-1$ the 2-from reads
\begin{align}
    &\omega^{(2)}_{\a>1}(t_1,t_2)=\frac{w^{(2)}_{\a>1}(t_1,t_2)}{t_1t_2(t_2-t_1)}dt_1\wedge dt_2, \label{temp1} \\
    &w^{(2)}_{\a>1}(t_1,t_2)=(-1)^{\a-1}z_1^{-t_1+n}z_2^{-t_2+n-1} \nonumber \\
    &\quad\times\frac{\Gamma(1-t_1)\Gamma(1+t_1)\Gamma(1-t_2)\Gamma(1+t_2)}{\Gamma(n+1-t_1)\Gamma(n+\a-t_2)}.
\end{align}
Due to presence of the $(t_2-t_1)$-term in the denominator of~(\ref{temp1}) we use the transformation law~(\ref{transformation_law}) with~$f_1=t_1$, $f_2=t_2(t_2-t_1)$, $g_1=t_1^2$, $g_2=t_2^2$, which yields
\begin{equation}
T=\lp\begin{array}{cc}t_1& 0\\ t_2 & 1 \end{array}\rp,
\end{equation}
\begin{align}
    \Res_{\bm{t}=0}\omega^{(2)}_{\a>1}(t_1,t_2) &= \Res_{\bm{t}=0}\frac{w^{(2)}_{\a>1}(t_1,t_2)}{t_1t_2^2}dt_1\wedge dt_2 \nonumber \\
    &=\pp_{t_2}w_{\a>1}^{(2)}(t_1,t_2)\Big\vert_{\tln{t_1\to0}{t_2\to0}}.
\end{align}
Therefore, the contribution is given by a single series
\begin{align} \label{MagentaGe}
H_\text{mag}^{\alpha\ge 1} = \frac{(-1)^{\a}}{z_2} \sum_{n=0}^{\infty}\frac{(z_1z_2)^n}{n!(n+\a-1)!} \big(\ln z_2 - \psi(n+\a)\big),
\end{align}
where $\psi(n+1) = H_n - \gamma$ is the digamma function, $H_n = \sum_{k=1}^n k^{-1}$ are the harmonic numbers, and $\gamma$ is the Euler constant.

\paragraph{``Magenta'' poles for $\alpha\le1$.}

When $\alpha\le1$, the ``magenta'' poles are at $(\a-n-1,\a-n)$, $n\in\bbZ_{\geq0}$,
\begin{align}
    &\omega_{\a<1}^{(2)}(t_1,t_2)=\frac{w_{\a<1}^{(2)}(t_1,t_2)}{t_1t_2(t_2-t_1)}dt_1\wedge dt_2, \\
    &w_{\a<1}^{(2)}(t_1,t_2)=(-1)^{1-\a}z_1^{-t_1-\a+n+1}z_2^{-t_2-\a+n} \nonumber \\
    &\quad\times\frac{\Gamma(1-t_1)\Gamma(1+t_1)}{\Gamma(n+2-\a-t_1)}\frac{\Gamma(1-t_2)\Gamma(1+t_2)}{\Gamma(n+1-t_2)}.
\end{align}
Once again, $f_1=t_1$, $f_2=t_2(t_2-t_1)$, $g_1=t_1^2$, $g_2=t_2^2$, $\det\h A=t_1$, so we have
\begin{align}
    \Res_{\bm{t}=0}\omega_{\a<1}^{(2)}(t_1,t_2)&=\Res_{\bm{t}=0}\frac{w_{\a<1}^{(2)}(t_1,t_2)}{t_1t_2^2}dt_1\wedge dt_2 \nonumber \\
    &=\pp_{t_2}w_{\a<1}^{(2)}(t_1,t_2)\Big\vert_{\tln{t_1\to0}{t_2\to0}},
\end{align}
so this family's contribution reads
\begin{equation} \label{MagentaLe}
H_\text{mag}^{\alpha\le 1} = \frac{(-1)^{\a}}{z_2} \sum_{n=0}^{\infty}\frac{(z_1z_2)^{n+1-\a}}{n!(n+1-\a)!} \big(\ln z_2 - \psi(n+1)\big).
\end{equation}

It is easy to see that, as expected, for the intermediate value $\alpha=1$, the expressions given by formulas \eqref{MagentaGe} and \eqref{MagentaLe} are identical. As a result, we obtain that the representation of the integral $H_1(z_1,z_2)$~\eqref{tildeE3} for $\mu=\nu=1$, $\a\in\bbZ$, is given by the sum of the regular $H_\text{blue}$ \eqref{BlueSeries} and the degenerate $H_\text{mag}$ contributions, the latter being given by formula \eqref{MagentaGe} for $\alpha\ge1$ and formula \eqref{MagentaLe} for $\alpha\le1$. Also, for $\alpha<1$, we must add the finite contribution \eqref{OrangePol}. This agrees with the answers obtained with the help of \texttt{MBConicHulls} Mathematica package~\cite{Ananthanarayan2021} for several fixed integer values of $\alpha$ both smaller and greater than 1.

These series can be utilized to illustrate the consistency of our procedure. In~\cite{BKW25a}, where we stated this answer without any derivation we used this series to show how the IR-divergent parts of basis kernels can be regularized. In particular, if we consider the basis kernel of the operator function $f(\h F)=\h F^{-1}\exp(-\tau \h F)$~\eqref{bbK_mn_from_bbK_0n}-\eqref{calKMellin} in the resonant case $\a=d/2-k\in\bbZ$, its the MB integral becomes ill-defined due to the contour pinch. Performing a dimensional regularization $d\mapsto d+2\epsilon$, $\alpha\mapsto\alpha+\epsilon$ and expanding in $\epsilon$ in the vicinity of this pinch one obtains expressions for finite $O(\epsilon^0)$ and divergent $O(\epsilon^{-1})$ parts of the basis kernel $\bbB_{\a}[F^{-1}\exp(-\tau F)|\s]$. Thanks to the answer \eqref{MagentaGe} and \eqref{MagentaLe}, we can interpret them as IR-finite and IR-divergent contributions to the series since, as a straightforward calculation shows, one has a direct correspondence between the dimensional regularization of the pinch and massive regularization, i.e. the $O(m^2)$-part of the expansion of the complete kernel~\eqref{h3mu1nu1} in $m^2\to0$ is identical to the finite part of the $\epsilon$-expansion of the basis kernel,
\begin{equation}
    \bbB_{\a+\epsilon}\!\lb \frac{e^{-\tau F}}{F}\Bigg|\s\rb\Bigg|^{\mathrm{fin}}_{\epsilon\to0}=\bbW_{\a}\!\lb \frac{e^{-\tau F}}{F}\Bigg|\s,m^2\rb\Bigg|^{\mathrm{fin}}_{m^2\to0}\!.
\end{equation}
Moreover, the divergent part of the $\epsilon$-expansion can be identified with the logarithmically divergent part of the $m^2\to0$ expansion of the complete kernel,
\begin{equation}
    \bbB_{\a+\epsilon}\!\lb \frac{e^{-\tau F}}{F}\Bigg|\s\rb\Bigg|^{\mathrm{div}}_{\epsilon\to0}=\bbW_{\a}\!\lb \frac{e^{-\tau F}}{F}\Bigg|\s,m^2\rb\Bigg|^{\mathrm{log\,div}}_{m^2\to0}\!,
\end{equation}
whereas power divergent terms of the complete massive kernel do not appear in the $\epsilon$-expansion at all, an expected property of dimensional regularization.

The aforementioned equivalence holds for the basis and complete kernels of any other operator functions and thus shows the consistency of our technique. This result allows us to interpret pinches of MB contour as IR-divergences of the corresponding kernels, which can be regularized either through dimensional regularization and analytic continuation or via introduction of the mass term to the operator, with the two regularization schemes producing agreeing results.  At the same time, this also serves as an additional verification of our approach, in which we avoid separate consideration of the resonant series, since they can be obtained from the non-resonant series by the appropriate limit procedure even in cases when the final answer is singular due to pinches.

\section{Conclusion}
\label{sec:conclusion}

This marks the end to the series of papers~\cite{BKWletter,BKW25a}, where we have developed a systematic method for obtaining off-diagonal expansions for integral kernels of functions of Laplace-type operators on curved background. The technique, based on the off-diagonal functoriality of the DeWitt expansion, yields asymptotic series in the basis kernels and the heat coefficients of the operator. Both basis and complete kernels are functions of the Synge world function, they are independent of the underlying geometric data and the exact nature of the Laplace-type operator itself. Hypergeometric nature of these functions makes it highly convenient to apply the formalism of $N$-fold Mellin-Barnes integrals to study them.

The developed method has direct applications to the development of heat kernel methods for nonminimal operators, necessary for studying various models with nonminimal wave operators. In general, it should be useful in many areas of theoretical physics and mathematics, in particular, the theory of pseudodifferential operators and its applications. It could also be of use in studying various theories, such as higher-derivative nonlocal gravities.

In the present paper we have obtained and analyzed series representations for basis and complete kernels for a family of Laplace-type operator family of functions $\exp(-\tau \hat F^{\nu})/(\hat F^{\mu}+\lambda)$ in both resonant and non-resonant cases. We have shown that the former case, despite being physically more relevant is, in some sense, less fundamental than the latter. Indeed, the series representation in the resonant case can be obtained by taking the appropriate limit of the non-resonant series in the parameter space. And, as illustrated in the last Sect.~\ref{ResonantSec} this limit procedure remains meaningful even if the resonant series bears singularities due to contour pinches in the corresponding Mellin-Barnes integrals.

Apart from the complete massive and basis kernels of the operator function $\exp(-\tau \hat F^{\nu})/(\hat F^{\mu}+\lambda)$, we have studied simpler operator functions such as the complex power $\hat F^{-\mu}$, the resolvent of the power $1/(\hat F^{\mu}+\lambda)$, the heat kernel of the power $\exp(-\tau\hat F^{\nu})$, and the hybrid kernel $\exp(-\tau \hat F^{\nu})/\hat F^\mu$. Using Mellin-Barnes representations for basis and complete massive kernels for these operator functions obtained in~\cite{BKW25a}, we have obtained series representations for these hypergeometric-type functions, which allowed us to explore and verify various limits which connect them. Apart from serving as a verification of the generalized functorial technique, calculating these limits allowed us to suggest direct physical meaning to the individual cluster series, which together comprise the complete series representation. We concluded that in general, the basis and complete kernels contain several types of terms, which in simple cases can be directly associated with the IR and UV limits.

Among the open problems related to the present generalized functorial technique we would like to specifically mention the computation of IR asymptotic series, at least in the simplest geometries. Another interesting direction for further research would be the complete classification of cluster series for multiple Mellin-Barnes integrals according to their physical meaning is of interest.

Among the possible generalizations of this approach we would like to mention two principal directions. Firstly, one can consider kernels of operators of non Laplace-type, be that nonminimal or higher-order ones. In both cases, the starting point of the functorial technique---the heat kernel---has a more complicated structure than a relatively simple DeWitt series, however, we believe that this generalization is possible. Another, much more difficult task is to obtain expansions for the kernels of functions of multiple operators. This problem is directly related to many areas of ongoing research including multiloop calculations on a curved background and is thus of great interest.

Regarding the technique of calculating series representations of $N$-fold Mellin-Barnes integrals, which we extensively use in the paper, we have encountered some subtleties with the method of~\cite{Ananthanarayan2021} for $N>2$. Although this does not affect any of the integrals we encountered, given the importance and scope of applications of MB integrals future work should focus on these issues.
\section*{Acknowledgments}
The work of A. O. B. and A. E. K. was supported by the
grant from the “BASIS” Foundation for the Advancement
of Theoretical Physics and Mathematics.
\appendix
\section{Euler gamma function and its properties\label{A}}

Euler gamma function is defined using the well-known integral
\begin{equation} \label{GammaDef}
\Gamma(z) = \int\limits_0^\infty t^{z-1} e^{-t} dt,
\end{equation}
absolutely converging for $\Re z>0$. Integrating by parts, we prove the key relation
\begin{equation} \label{GammaRec}
\Gamma(z+1) = z\Gamma(z),
\end{equation}
from which it follows that the gamma function is a generalization of the factorial $\Gamma(n+1) = n!$, since~\eqref{GammaDef} yields $\Gamma(1)=1$. On the other hand, using the relation \eqref{GammaRec} the gamma function can be analytically extended to the domain $\Re z\le 0$ to a meromorphic function defined on the entire complex plane, except for simple poles at the points $z=-n$ with residues
\begin{equation} \label{GammaRes}
\Res_{z=-n} \Gamma(z) = \frac{(-1)^n}{n!}.
\end{equation}

Important properties of the gamma function are the reflection formula
\begin{equation} \label{ReflectionFormula}
\Gamma(z)\Gamma(1-z) = \frac{\pi}{\sin(\pi z)},
\end{equation}
and the multiplication theorem
\begin{equation}
\prod\limits_{k=0}^{m-1} \Gamma\big(z + \tfrac{k}{m}\big) = (2\pi)^\frac{m-1}{2} m^{\frac{1}{2} - mz} \Gamma(mz),
\label{mult_thrm}
\end{equation}
which at $m=2$ turns into the well-known Legendre duplication formula.

The well-known beta integral is expressed in a simple way through the gamma function:
\begin{equation} \label{BetaFunction}
B(x,y) = \int\limits_0^1 t^{x-1}(1-t)^{y-1} dt = \frac{\Gamma(x)\Gamma(y)}{\Gamma(x+y)}.
\end{equation}
Another relation, often used in physics (in calculating Feynman integrals), has the form:
\begin{equation} \label{PowerOfSum}
(x+y)^{-\alpha} = \frac{x^{-\alpha}}{\Gamma(\alpha)} \int\limits_C \frac{ds}{2\pi i}\, \Gamma(-s)\Gamma(\alpha+s) \left(\frac{x}{y}\right)^{-s},
\end{equation}
where the integration contour $C$ separates leftward- and rightward-running poles.

A universal tool for analyzing asymptotic behavior of gamma function is the asymptotic expansion
\begin{align}
\ln\Gamma(a + s) &\approx \ln\sqrt{2\pi} + \left(a + s - \tfrac{1}{2}\right)\ln s - s \nonumber \\
&+ \sum\limits_{k=1}^\infty \frac{(-1)^{k+1}}{k(k+1)}B_{k+1}(a) s^{-k},
\end{align}
for $|\,s|\,\to\infty$ and $|\,\arg s|\,<\pi-\varepsilon$ (here $B_k$ are Bernoulli polynomials). Dropping the sum in the second line of this expansion yields the famous Stirling formula:
\begin{equation} \label{Stirling}
\Gamma(a + s) = \sqrt{2\pi}\, s^{a + s - \tfrac{1}{2}}\, e^{-s} \left(1 + O(s^{-1})\right).
\end{equation}

\bibliography{Wachowski2603_ak}
\end{document}